\pdfoutput=1
\newcommand*{\ATLASLATEXPATH}{latex/}
\documentclass[USenglish, texlive=2016, cernpreprint]{\ATLASLATEXPATH atlasdoc}

\usepackage[subfigure=true, subfig=false]{\ATLASLATEXPATH atlaspackage}
\usepackage{\ATLASLATEXPATH atlasbiblatex}
\usepackage{\ATLASLATEXPATH atlascontribute}
\usepackage[BSM,hepparticle,hepprocess, jetetmiss,unit]{\ATLASLATEXPATH atlasphysics}
\usepackage{units}
\usepackage{SUSY-2016-30-PAPER-defs}

\graphicspath{{logos/}{figures/}}

\AtlasTitle{Search for squarks and gluinos in final states with hadronically decaying $\tau$-leptons, jets, and missing transverse momentum 
using $pp$ collisions at $\sqrt{s}$~=~13~\tev\ with the ATLAS detector}

\AtlasAbstract{%
A search for supersymmetry in events with large missing transverse momentum, jets, and at least one hadronically decaying $\tau$-lepton is presented. 
Two exclusive final states with either exactly one or at least two $\tau$-leptons are considered.
The analysis is based on proton--proton collisions at $\sqrt{s}=13$~\tev\ corresponding to 
an integrated luminosity of 36.1~fb$^{-1}$ 
delivered by the Large Hadron Collider and recorded by the ATLAS detector in 2015 and 2016.
No significant excess is observed over the Standard Model expectation. 
At 95\% confidence level, model-independent upper limits on the cross section are set and exclusion limits are provided for two signal scenarios:
a simplified model of gluino pair production with $\tau$-rich cascade decays, and a model with gauge-mediated supersymmetry breaking (GMSB).
In the simplified model, gluino masses up to 2000~\gev\ are excluded for low values of the mass of the lightest supersymmetric particle (LSP), while LSP masses up to 1000~\gev\ are excluded 
for gluino masses around 1400~\gev. 
In the GMSB model, values of the supersymmetry-breaking scale are excluded below 110~\tev\ for all values of $\tan\beta$ in the range $2\leq\tan\beta\leq60$, and below 120~\tev\ for $\tan\beta>30$.
}

\author{The ATLAS Collaboration}

\AtlasRefCode{SUSY-2016-30}

\PreprintIdNumber{CERN-EP-2018-185}

\arXivId{1808.06358v2}

\AtlasJournalRef{\PRD 99 (2019) 012009}
\AtlasDOI{10.1103/PhysRevD.99.012009}

\hypersetup{pdftitle={ATLAS document},pdfauthor={The ATLAS Collaboration}}

\begin{document}

\maketitle

\tableofcontents

\section{Introduction}
\label{sec:intro}

Supersymmetry (SUSY)~\cite{Golfand:1971iw,Volkov:1973ix,Wess:1974tw,Wess:1974jb,Ferrara:1974pu,Salam:1974ig}
introduces a symmetry between fermions and bosons, resulting in a SUSY partner (sparticle) for each Standard Model 
(SM) particle with identical quantum numbers except for a difference of half a unit of spin. Squarks ($\tilde q$), gluinos ($\tilde g$), charged sleptons ($\tilde \ell$), and sneutrinos ($\tilde \nu$) are the superpartners of the quarks, gluons, charged leptons, and neutrinos, respectively.
The SUSY partners of the gauge and Higgs bosons are called gauginos and higgsinos, respectively.
The charged electroweak gaugino and higgsino states mix to form charginos ($\tilde\chi_{i}^{\pm}$, $i$ = 1,2), 
and the neutral states mix to form neutralinos ($\tilde\chi_{j}^{0}$, $j$ = 1,2,3,4).
Finally, the gravitino ($\tilde G$) is the SUSY partner of the graviton.
As no supersymmetric particle has been observed, SUSY must be a broken symmetry. 
To avoid large violations of baryon- or lepton-number conservation, $R$-parity~\cite{Farrar:1978xj} 
conservation is often assumed. In this case, sparticles are produced in pairs and decay through cascades involving SM particles and other 
sparticles until the lightest sparticle (LSP), which is stable, is produced.

Final states with $\tau$-leptons are of particular interest in SUSY searches, although they are experimentally challenging.
Light sleptons could play a role in the co-annihilation of neutralinos in the early universe, and models with light $\tau$-sleptons are consistent with constraints on dark matter consisting of weakly interacting massive particles~\cite{Vasquez:2011,Bagnaschi:2017tru,Bechtle:2015nua}. Furthermore, should SUSY or any other physics beyond the 
Standard Model (BSM) be discovered in leptonic final states, independent studies of all three lepton flavors are necessary to investigate the coupling 
structure of the new physics, especially with regard to lepton universality. 

In this article, an inclusive search for squarks and gluinos produced via the strong interaction in events with jets (collimated sprays of particles from the hadronization of quarks and gluons), at least one 
hadronically decaying $\tau$-lepton, and large missing transverse momentum is presented.
Two SUSY models are considered: a simplified model~\cite{Alwall:2008ve,Alwall:2008ag,Alves:2011wf} of gluino pair production and a model of gauge-mediated SUSY breaking (GMSB)~\cite{Dine:1981gu,AlvarezGaume:1981wy,Nappi:1982hm}.
If squarks and gluinos are within the reach of the Large Hadron Collider (LHC), their production may be among the dominant SUSY processes. 
Final states with exactly one $\tau$-lepton (\onetau) or at least two $\tau$-leptons (\twotau) provide complementary acceptance to SUSY signals.
These two channels are optimized separately and the results are statistically combined.
Models with a small mass splitting between gluinos or squarks and the LSP, producing soft $\tau$-leptons in the final state, are best covered by the \onetau channel.
Models with a heavy LSP, producing signatures with low missing transverse momentum, are more easily probed by the \twotau channel due to the lower SM background.
For models with a large mass splitting, both channels provide sensitivity.

The analysis is performed using proton--proton ($pp$) collision data at a center-of-mass energy of 
$\sqrt{s}=13 \tev$ corresponding to an integrated luminosity of $36.1~\ifb$, recorded with the ATLAS detector 
at the LHC in 2015 and 2016. For both SUSY models, the exclusion limits obtained significantly improve upon the previous ATLAS results.
Besides the increase in the integrated luminosity, the results benefit from an improved analysis and statistical treatment.
Previous searches in the same final state have been reported by the ATLAS~\cite{SUSY-2012-07,SUSY-2013-10,SUSY-2016-01} and 
CMS~\cite{CMS-SUS-12-004} collaborations.

In GMSB models, SUSY breaking is communicated from a hidden sector to the visible sector by a set of messenger fields that share the gauge 
interactions of the SM. SUSY is spontaneously broken in the messenger sector, leading to massive, non-degenerate messenger fields.
The free parameters of GMSB models are the SUSY-breaking mass scale in the messenger sector ($\Lambda$), the messenger mass scale ($M_\mathrm{mes}$), the number of messenger multiplets ($N_5$) of the $\mathbf{5}+\mathbf{\bar 5}$ representation of $\text{SU}(5)$, the ratio of the two Higgs-doublet vacuum expectation values at the electroweak scale ($\tan\beta$), the sign of the Higgsino mass term in the superpotential ($\mathrm{ sign}(\mu)=\pm1$), and a gravitino-mass scale factor ($C_\mathrm{grav}$).
Details of the GMSB scenarios studied herein can be found in Ref.~\cite{SUSY-2016-01}.

As in previous ATLAS searches, the GMSB model is probed as a function of $\Lambda$ and $\tan\beta$, while the other 
parameters are set to $M_\mathrm{mes} =250$~\tev, $N_{\mathrm 5}=3$, $\mathrm{sign}(\mu)=1$, and $C_\mathrm{grav}=1$.
The choice of $\tan\beta$ influences the nature of the NLSP. 
For large values of $\tan\beta$, the NLSP is the $\PSgti$\footnote{The $\PSgti$ is the lighter of the two $\tau$-slepton mass eigenstates, which results from the mixing of the superpartners of the left- and right-handed $\tau$-leptons ($\tilde{\tau}_L$, $\tilde{\tau}_R$).} while 
for lower $\tan\beta$ values, the $\PSgti$ and the superpartners of the right-handed electron and muon 
($\PSe_{\text{R}},\PSgm_{\text{R}}$) are almost degenerate in mass.
The production of squark pairs dominates at high values of $\Lambda$, with a subdominant contribution from squark--gluino production. A typical GMSB signal process is 
displayed in Figure~\ref{fig:GMSBGraph}. The value of $C_\mathrm{grav}$ corresponds to prompt decays of the NLSP.

Although minimal GMSB cannot easily accommodate a Higgs boson with mass of approximately 125~\gev, various extensions exist (see, e.g., Refs.~\cite{GGM,nGM}) that remedy these shortcomings while preserving very similar signatures, in particular the natures of the LSP and the NLSP.

\begin{figure}[tb]
  \centering
  \subfigure[GMSB model]{
    \includegraphics[width=.42\textwidth]{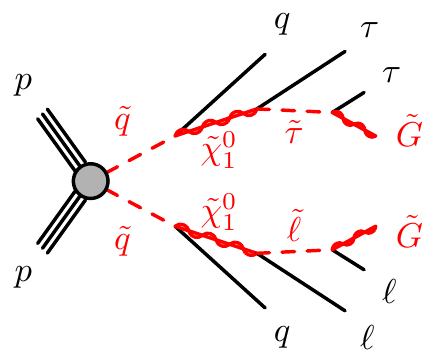}
    \label{fig:GMSBGraph}
  }
  \hspace{5mm}
  \subfigure[Simplified model]{
    \includegraphics[width=.42\textwidth]{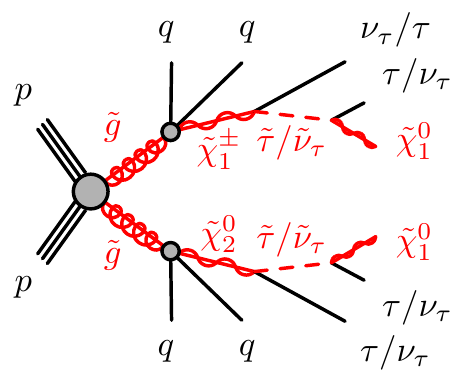}
    \label{fig:SimplifiedModelGraph}
  }
 \caption{Example processes of (a) the GMSB model and (b) the simplified model of gluino pair production leading to final states with $\tau$-leptons, jets and missing transverse momentum.}
  \label{fig:FeynmanDiagram}
\end{figure}

The simplified model of gluino pair production is inspired by generic models such as the $R$-parity-conserving phenomenological 
MSSM~\cite{Djouadi:1998di,Berger:2008cq} with dominant gluino pair production, light $\tilde\tau_1$, and a 
$\tilde\chi_1^0$ LSP. Gluinos are assumed to undergo a two-step cascade decay leading to $\tau$-rich final states, 
as shown in Figure~\ref{fig:SimplifiedModelGraph}.
The two free parameters of the model are the masses of the gluino ($m_{\tilde g}$) and the LSP ($m_{\tilde\chi_1^0}$).
Assumptions are made about the masses of other sparticles, namely the $\tilde\tau_1$ and $\tilde\nu_\tau$ are mass-degenerate, 
and the $\tilde\chi_2^0$ and $\tilde\chi_1^\pm$ are also mass-degenerate, with
\begin{equation*}
m_{\tilde\chi_1^\pm} = m_{\tilde\chi_2^0}= \frac{1}{2}(m_{\tilde g} + m_{\tilde\chi_1^0}) \; , \quad
m_{\tilde\tau_1}=m_{\tilde\nu_\tau}=\frac{1}{2}(m_{\tilde\chi_1^\pm} + m_{\tilde\chi_1^0}).
\end{equation*}
Gluinos are assumed to decay into $\tilde{\chi}^{\pm}_{1} q \bar q'$ and $\tilde{\chi}^0_{2} q \bar q$ 
with equal branching ratios, where $q,q'$ denote generic first- and second-generation quarks.
The neutralino $\tilde{\chi}^0_{2}$ is assumed to decay into $ \stau \tau$ or $ \tilde \nu_{\tau} \nu_{\tau}$ 
with equal probability, while the chargino $\tilde{\chi}^{\pm}_{1}$ is assumed to decay into 
$ \tilde \nu_{\tau}\tau$ or $ \stau\nu_{\tau}$ with equal probability.
In the last step of the decay chain, $\stau $ and $\tilde \nu_{\tau}$ are assumed to decay into 
$\tau \tilde{\chi}^0_{1}$ and $ \nu_\tau \tilde{\chi}^0_{1}$, respectively. 
All other SUSY particles are kinematically decoupled. 
The topology of signal events depends on the mass-splitting between the gluino and the LSP.  
The sparticle decay widths are assumed to be small compared to sparticle masses, such that they play no role in the kinematics.

\section{ATLAS detector}
\label{sec:detector}

The ATLAS experiment is described in detail in Ref.~\cite{PERF-2007-01}.
It is a multipurpose detector with a forward--backward symmetric cylindrical
geometry and a solid angle\footnote{ATLAS uses a right-handed coordinate
system with its origin at the nominal interaction point in the
center of the detector and the $z$-axis along the beam pipe. The
$x$-axis points from the interaction point to the center of the LHC ring and the
$y$-axis points upward. Cylindrical coordinates $(r,\phi)$ are used
in the transverse plane, $\phi$ being the azimuthal angle around the
beam pipe. The pseudorapidity is defined in terms of the polar angle
$\theta$ as $\eta=-\ln\tan(\theta/2)$. Rapidity is defined as $y = 0.5\ln[\nicefrac{(E+p_z)}{(E-p_z)}]$ where $E$ denotes the energy and $p_z$ represents the momentum component along the $z$-axis.} coverage of nearly 4$\pi$.

The inner tracking detector (ID), covering the region $|\eta|<2.5$, consists
of a silicon pixel detector, a silicon microstrip detector, and a
transition radiation tracker. The innermost layer of the pixel detector, the insertable B-layer~\cite{IBL}, was installed between Run 1 and Run 2 of the LHC.
The inner detector is surrounded by a thin superconducting solenoid providing a \unit[2]{T} magnetic field, and by
a finely segmented lead/liquid-argon (LAr) electromagnetic calorimeter covering the region $|\eta|<3.2$.
A steel/scintillator-tile hadronic calorimeter provides coverage in
the central region $|\eta|<1.7$. The endcap and
forward regions, covering the pseudorapidity range $1.5<|\eta|<4.9$, are instrumented with electromagnetic and hadronic
LAr calorimeters, with steel, copper or tungsten as the absorber material.
A muon spectrometer system incorporating large
superconducting toroidal air-core magnets surrounds the calorimeters. Three layers of precision wire chambers 
provide muon tracking coverage in the range $|\eta|<2.7$, while dedicated fast chambers are used for triggering in the region $|\eta|<2.4$.

The trigger system is composed of two stages~\cite{TRIG-2016-01}.
The level-1 trigger, implemented with custom hardware, uses information from calorimeters and muon chambers to reduce the event rate from 40~MHz to a maximum of 100~kHz.
The high-level trigger reduces the data acquisition rate to about 1~kHz. 
It is software based and runs reconstruction algorithms similar to those used in the offline reconstruction.

\section{Data and simulated event samples}
\label{sec:samples}

The data used in this analysis consist of $pp$ collisions at a center-of-mass energy of $\sqrt{s}=13$~\tev\ delivered by the LHC with a 25~ns bunch spacing and recorded by the ATLAS detector in 2015 and 2016. The average number of $pp$ interactions per bunch crossing, $\langle\mu\rangle$, was 13.4 in 2015 and 25.1 in 2016.
Data quality requirements are applied to ensure that all subdetectors were operating normally, and that LHC beams were in stable collision mode.
The integrated luminosity of the resulting data set is $36.1$~\ifb. 

Monte Carlo (MC) simulations are used to model both the SUSY signals and SM backgrounds, except for multijet production, which is evaluated from data.
Soft $pp$ interactions (pileup) were included in the simulation using the {\sc Pythia}~8.186~\cite{Sjostrand:2014zea} generator with the A2~\cite{ATL-PHYS-PUB-2012-003} set of tuned parameters (minimum-bias tune) and MSTW2008LO~\cite{Martin:2009iq} parton distribution function (PDF) set. Generated events were reweighted such that the $\langle\mu\rangle$ distribution of the simulation matches the one in data.
For SM background samples, the interactions between particles and the detector material were simulated~\cite{SOFT-2010-01} using {\sc Geant4}~\cite{Agostinelli:2002hh} and a detailed description of the ATLAS detector.
For signal samples, a parameterized fast simulation was used to describe the energy deposits in the calorimeters~\cite{ATL-SOFT-PUB-2014-01}.

The $W$+jets and $Z$+jets ($V$+jets) processes were simulated with the {\sc Sherpa}~\cite{Gleisberg:2008ta} generator using version~2.2.1.
Matrix elements (ME) were calculated for up to two partons at next-to-leading order (NLO) and up to four additional partons at leading order (LO) in perturbative QCD using the {\sc OpenLoops}~\cite{Cascioli:2011va} and {\sc Comix}~\cite{Gleisberg:2008fv} ME generators, respectively.
The phase space merging between the {\sc Sherpa} parton shower (PS)~\cite{Schumann:2007mg} and MEs followed the ME+PS@NLO prescription~\cite{Hoeche:2012yf}. 
The NNPDF3.0nnlo~\cite{nnpdf30} PDF set was used in conjunction with dedicated parton-shower tuning. 
The inclusive cross sections were normalized to a next-to-next-to-leading-order (NNLO) calculation~\cite{Catani:2009sm} in perturbative QCD based on the {\sc FEWZ} program~\cite{Anastasiou:2003ds}.
An additional $W(\tau\nu)$ sample is used for evaluating systematic uncertainties; this was generated with {\sc MG5\_aMC@NLO} v2.2.3~\cite{Alwall:2014hca} interfaced to {\sc Pythia} 8.186 with the A14 tune~\cite{ATL-PHYS-PUB-2014-021} for the modeling of the PS, hadronization, and underlying event. The ME calculation was performed at tree level and includes the emission of up to four additional partons. The PDF set used for the generation was NNPDF23LO~\cite{Ball:2013hta}.

For the simulation of $t\bar t$ events, the {\sc Powheg-Box} v2~\cite{Alioli:2010xd} generator was used with the CT10~\cite{Lai:2010vv} PDF set for the ME calculation.
Electroweak single-top-quark production in the $s$-channel, $t$-channel and $Wt$ final state was generated using {\sc Powheg-Box} v1. 
The PS, hadronization, and underlying event were simulated using {\sc Pythia}~6.428~\cite{Sjostrand:2006za} with the CTEQ6L1~\cite{Pumplin:2002vw} PDF set and the corresponding Perugia~2012 tune~\cite{Skands:2010ak}. 
Cross sections were calculated at NNLO in perturbative QCD with resummation of next-to-next-to-leading-logarithm (NNLL) soft gluon terms using the {\sc Top++} 2.0 program~\cite{Czakon:2011xx}.

Diboson production was simulated using {\sc Sherpa} 2.2.1 and~2.2.2 with the NNPDF3.0nnlo PDF set.
Processes with fully leptonic final states were calculated with up to one parton for the $4\ell$, $2\ell+2\nu$ samples or no parton for the $3\ell+1\nu$ samples at NLO and up to three additional partons at LO. 
Diboson processes with one of the bosons decaying hadronically and the other leptonically were simulated with up to one parton for the $ZZ$ or no parton for the $WW$ and $WZ$ samples at NLO, and up to three additional partons at LO. 
The cross section provided by the generator is used for these samples.

The simplified-model signal samples were generated using {\sc MG5\_aMC@NLO} v2.2.3 interfaced to {\sc Pythia} 8.186 with the A14 tune. The ME calculation was performed at tree level and includes the emission of up to two additional partons. The PDF set used for the generation was NNPDF23LO. The ME--PS matching was performed using the CKKW-L prescription, with a matching scale set to one quarter of the gluino mass. 
The GMSB signal samples were generated with the \Herwigpp~2.7.1~\cite{Bahr:2008pv} generator, with CTEQ6L1 PDFs and the UE-EE-5-CTEQ6L1 tune~\cite{Gieseke:2012ft}, 
using input files generated in the SLHA format with the {\sc SPheno} v3.1.12~\cite{Porod:2011nf} program.
The PS evolution was performed using an algorithm described in Refs.~\cite{Bahr:2008pv,Marchesini:1983bm,Marchesini:1987cf,Gieseke:2003rz}.
Signal cross sections were calculated to next-to-leading order in the strong coupling constant, adding the resummation of soft gluon emission at next-to-leading-logarithm accuracy (NLO+NLL)~\cite{Beenakker:1996ch,Kulesza:2008jb,Kulesza:2009kq,Beenakker:2009ha,Beenakker:2011fu}. The nominal cross section and its uncertainty were taken from an envelope of cross-section predictions using different PDF sets and factorization and renormalization scales, as described in Ref.~\cite{Borschensky:2014cia}.

\section{Event reconstruction}
\label{sec:reconstruction}

This search is based on final states with jets, hadronically decaying $\tau$-leptons, and missing transverse momentum. In addition, muons and $b$-tagged jets are used for background modeling studies, while electrons are only used for the missing transverse momentum calculation.

Interaction vertices are reconstructed using inner-detector tracks with transverse momentum $\pt>400 \mev$~\cite{ATL-PHYS-PUB-2015-026}.
Primary vertex candidates are required to have at least two associated tracks, and the candidate with the largest $\sum\pt^2$ is defined as the primary vertex. Events without a reconstructed primary vertex are rejected.

Jets are reconstructed using the anti-$k_t$ clustering algorithm~\cite{Cacciari:2008gp,fastjet} with a distance parameter $R=0.4$.
Clusters of calorimeter cells~\cite{PERF-2014-07}, calibrated at the electromagnetic energy scale, are used as input.
The jet energy is calibrated using a set of global sequential calibrations~\cite{ATLAS-CONF-2015-002,ATL-PHYS-PUB-2015-015}. 
Jets are required to have $\pt > 20 \gev$ and $|\eta| < 2.8$.
A jet-vertex-tagging algorithm~\cite{PERF-2014-03} is used to discriminate hard-interaction jets 
from pileup jets for jets with $|\eta|<2.4$ and $\pt < 60 \gev$.
Events with jets originating from cosmic rays, beam background or detector noise are rejected~\cite{ATLAS-CONF-2015-029}. 
Jets containing $b$-hadrons ($b$-jets) are identified using a multivariate algorithm exploiting the long lifetime, high decay
multiplicity, hard fragmentation, and large mass of $b$-hadrons~\cite{ATL-PHYS-PUB-2015-022}. The $b$-tagging algorithm identifies $b$-jets with an efficiency of approximately 70\% in simulated $t\bar t$ events.
The rejection factors for $c$-jets, hadronically decaying $\tau$-leptons, and light-quark or gluon jets are approximately 8, 26 and 440, respectively~\cite{ATL-PHYS-PUB-2015-039}.

Muon candidates are reconstructed in the region $|\eta| < 2.5$ from muon spectrometer tracks matching
ID tracks. Muons are required to have $\pt > 10 \gev$ and pass {\it medium} identification requirements~\cite{Aad:2016jkr}, based on the number of hits in the ID and muon spectrometer, and the compatibility of the charge-to-momentum ratios measured in the two detector systems. Events containing poorly reconstructed muons or cosmic-ray muon candidates are rejected.
Details of the electron reconstruction are given in Refs.~\cite{PERF-2013-05,ATL-PHYS-PUB-2015-041}.

Hadronically decaying $\tau$-leptons are reconstructed~\cite{ATL-PHYS-PUB-2015-045} from anti-$k_t$ jets within $|\eta|<2.5$ calibrated with a local cluster weighting technique~\cite{PERF-2011-03}.
The $\tau$-lepton candidates are built from clusters of calorimeter cells within a cone of size $\delReta \equiv \sqrt{\smash[b]{(\Delta\eta)^2+(\Delta\phi)^2}}=0.2$ centered on the jet axis.
A boosted regression tree is used to calibrate the energy of reconstructed $\tau$-leptons.
It exploits shower-shape information from the calorimeter, the track multiplicity, the amount of pileup, and information from particle-flow reconstruction~\cite{PERF-2014-06} that aims to identify charged and neutral hadrons from the $\tau$-lepton decay.
The $\tau$-leptons are required to have $\pt>20 \gev$, and candidates reconstructed within the transition region between the barrel and endcap calorimeters, $1.37 < |\eta| <1.52$, are discarded.
The $\tau$-leptons are required to have either one or three associated tracks, with a charge sum of $\pm 1$. 
A boosted-decision-tree discriminant is used to separate jets from $\tau$-leptons.
It relies on track variables from the inner detector as well as shower-shape variables from the calorimeters.
The analysis makes use of {\it loose} and {\it medium} $\tau$-leptons, corresponding to identification efficiencies of 60\% and 55\%, respectively, for one-track $\tau$-leptons and 50\% and 40\%, respectively, for three-track $\tau$-leptons.
Electrons reconstructed as one-track $\tau$-leptons are rejected by imposing a \pt- and $|\eta|$-dependent requirement on the likelihood identification variable of the electron, which provides a constant efficiency of 95\% for real $\tau$-leptons, with a rejection factor for electrons ranging from 30 to 150 depending on the $|\eta|$ region.
Like for jets, events with $\tau$-lepton candidates close to inactive calorimeter regions are rejected.

The missing transverse momentum vector \ptmissv, whose magnitude is denoted by \met, is defined as the negative vector sum of the transverse momenta of all identified and calibrated physics objects (electrons, muons, jets, and $\tau$-leptons) and an additional soft term.
The soft term is constructed from all the tracks with $\pt>400\mev$ which originate from the primary vertex but are not associated with any physics object. 
This track-based definition makes the soft term largely insensitive to pileup~\cite{ATL-PHYS-PUB-2015-027}.

After the reconstruction, an {\it overlap-removal} procedure is applied to remove ambiguities in case the same object is reconstructed by different algorithms. The successive steps of this procedure are summarized in Table~\ref{tab:OverlapRemoval}, where the overlap of reconstructed objects is defined in terms of the distance between objects $\delRy \equiv \sqrt{\smash[b]{(\Delta y)^2+(\Delta\phi)^2}}$. 
First, \textit{loose} $\tau$ candidates are discarded if they overlap with an electron or muon (steps 1 and 2). If an electron and a muon are reconstructed using the same inner-detector track, the electron is discarded (step 3). 
For overlapping light leptons (electrons and muons) and jets, the jet is kept in cases where the lepton is likely to result from a heavy-flavor hadron decay within the jet, otherwise the lepton is kept (steps 4--7).
Finally, if a jet is also reconstructed as a \textit{loose} $\tau$-lepton, the jet is discarded (step 8).

\begin{table}[htb]
 \caption{Overview of the successive steps in the overlap-removal procedure.}
  \centering
  \begin{tabular}{clll}
    \toprule
         & Object discarded            & Object kept      & Matching condition                  \\
    \midrule
    1.   & loose $\tau$                 &  electron            &  $\delRy < 0.2$         \\ 
    2.   & loose $\tau$                 &  muon                & $\delRy < 0.2$           \\ 
     3.     &  electron         &  muon                & shared inner-detector track    \\ 
    4.   & jet                       &  electron            & $\delRy < 0.2$ and jet not $b$-tagged        \\
     5.  &  electron         & jet                          & $\delRy < 0.4$        \\ 
    6.   &  jet             & muon                          & $\delRy < 0.2$, jet with $\leq2$ tracks and not $b$-tagged       \\
     7.  &  muon             & jet                          & $\delRy < 0.4$        \\ 
    8.   & jet                       & loose $\tau$                    & $\delRy < 0.2$ \\ 
    \bottomrule
  \end{tabular}
   \label{tab:OverlapRemoval}
\end{table}

\section{Event selection}
\label{sec:selection}

A preselection common to the \onetau and \twotau channels is applied. 
Events are required to pass the missing transverse momentum trigger with the lowest threshold and no bandwidth limitation.
To select a phase space where the trigger is fully efficient, the offline selection requires $\met > 180 \gev $ and a leading jet with $\pt > 120 \gev$.
Furthermore, an additional jet with $\pt > 25 \gev$ is required.
The two leading jets are required to be separated from \ptmissv by at least $0.4$ in $\phi$, to reject multijet background where large \met can arise from jet energy mismeasurements. 
The \onetau channel requires exactly one \textit{medium} $\tau$-lepton while the \twotau channel requires at least two \textit{medium} $\tau$-lepton. The preselection is summarized in Table~\ref{tab:presel}.

\begin{table}[tb]
  \caption{Summary of the preselection criteria applied in the \onetau and \twotau channels. $\njet$ and $\ntau$ are the number of jets and $\tau$-leptons respectively; other variables are defined in the text.}
  \centering
  \begin{tabular}{l|c|c}
    \toprule
    Subject of selection  & \onetau channel & \twotau channel\\
    \midrule
 Trigger & \multicolumn{2}{c}{$\met > \SI{180}{\giga\electronvolt},\;\LjetPt > \SI{120}{\giga\electronvolt}$}\\
\midrule
   Jets & \multicolumn{2}{c}{$\njet \ge 2,\;\SLjetPt > \SI{25}{\giga\electronvolt}$}\\
    \midrule
    Multijet events & \multicolumn{2}{c}{$\dphijets > 0.4$}\\
    \midrule
    $\tau$-leptons & $\ntau = 1$ & $\ntau \geq 2$\\
    \bottomrule
  \end{tabular}
\label{tab:presel}
\end{table}

To isolate signatures of potential SUSY processes from known SM background, additional kinematic variables are utilized:
\begin{itemize}
\renewcommand\labelitemi{---}
\item{The transverse mass of the system formed by \ptmissv and the momentum ${\boldsymbol p}$ of a reconstructed object,
  \begin{equation*}
  \mT \equiv m_\mathrm{T}({\boldsymbol p},\ptmissv)=\sqrt{2\pt\met(1-\cos\Delta\phi({\boldsymbol p},\ptmissv))}~,
  \end{equation*} 
  where $\Delta\phi({\boldsymbol p},\ptmissv)$ denotes the azimuthal angle between the momentum of the reconstructed object and the missing transverse momentum.
  For events where a lepton $\ell$ and the missing transverse momentum both originate from a $W(\ell\nu)$ decay, the $\mTl$ distribution exhibits a Jacobian peak at the $W$ boson mass.
  The transverse mass of various objects is used in this analysis, most notably the transverse mass of the reconstructed $\tau$-lepton.}
\item{The $\mttwo^{\tau\tau}$ variable~\cite{Lester:1999tx,Lester:2014yga}, also called {\it stransverse mass}, computed as
\begin{equation*}
\mttwo^{\tau\tau} = \min\limits_{{\boldsymbol{p}_{\mathrm{T}}^{a}+ \boldsymbol{p}_{\mathrm{T}}^{b} = \ptmissv }}\left(\max\left[m_\mathrm{T}(\boldsymbol{p}^{\tau_1},\boldsymbol{p}_{\mathrm{T}}^{a}) , m_{\mathrm{T}}(\boldsymbol{p}^{\tau_2},\boldsymbol{p}_{\mathrm{T}}^{b})   \right]\right)~,
\label{eq:mttwo}
\end{equation*}
where $(a,b)$ refers to two invisible particles that are assumed to be produced with transverse momentum vectors $\boldsymbol{p}_\mathrm{T}^{~a,b} $. In this calculation, $(a,b)$ are assumed to be massless.
The $\mttwo^{\tau\tau}$ distribution has a kinematic endpoint for processes where massive particles are pair-produced, each particle decaying into a $\tau$-lepton and an undetected particle.
When more than two $\tau$-leptons are produced in a decay chain, there is no way to a priori select the pair leading to the desired characteristic.
Therefore, $\mttwo^{\tau\tau}$ is calculated using all possible $\tau$-lepton pairs and the largest value is chosen.}
\item{The scalar sum of the transverse momenta of all $\tau$-leptons and jets,  
    $\HT=\sum\limits_i \pt^{\mathrm{\tau}_i}+ \sum\limits_j \pt^{\mathrm{jet}_j}$~.}
\end{itemize}

Figure~\ref{fig:preselection} shows examples of kinematic distributions after the preselection and after applying background normalization factors as described in section~\ref{sec:background}.
The dominant backgrounds in the \onetau channel are $t\bar t$ production and $W(\tau\nu)$+jets events, with subdominant contributions from $Z(\nu\nu)$+jets and $Z(\tau\tau)$+jets. 
In the \twotau channel, the spectrum is dominated by $t\bar t$, $W(\tau\nu)$+jets and $Z(\tau\tau)$+jets events. The multijet background does not contribute significantly while contributions from the diboson background are only relevant at high values of \mTTT. 

\begin{figure}[htb]
  \centering
  \subfigure[]{\includegraphics[width=.48\textwidth]{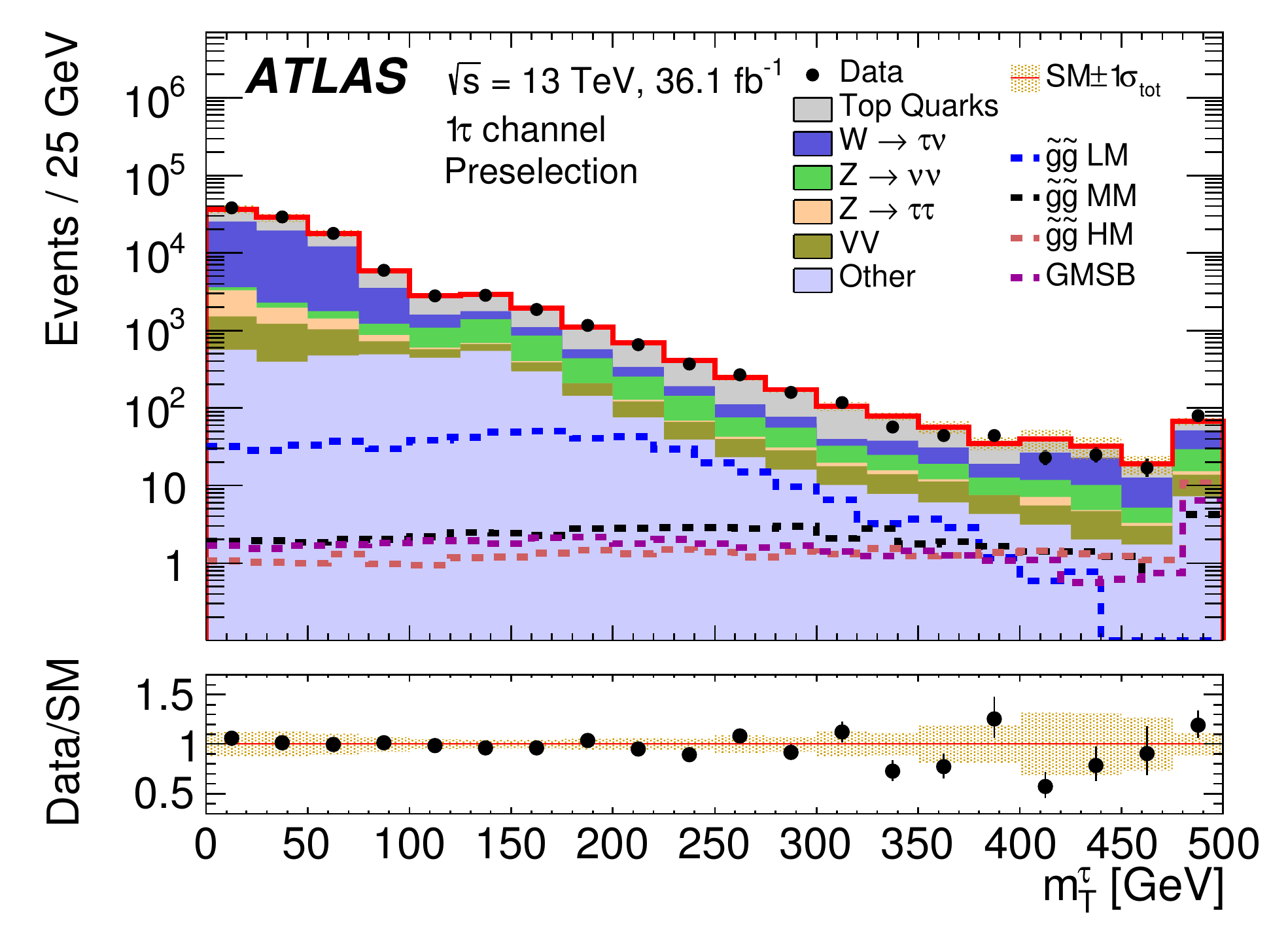}
                 }
  \subfigure[]{\includegraphics[width=.48\textwidth]{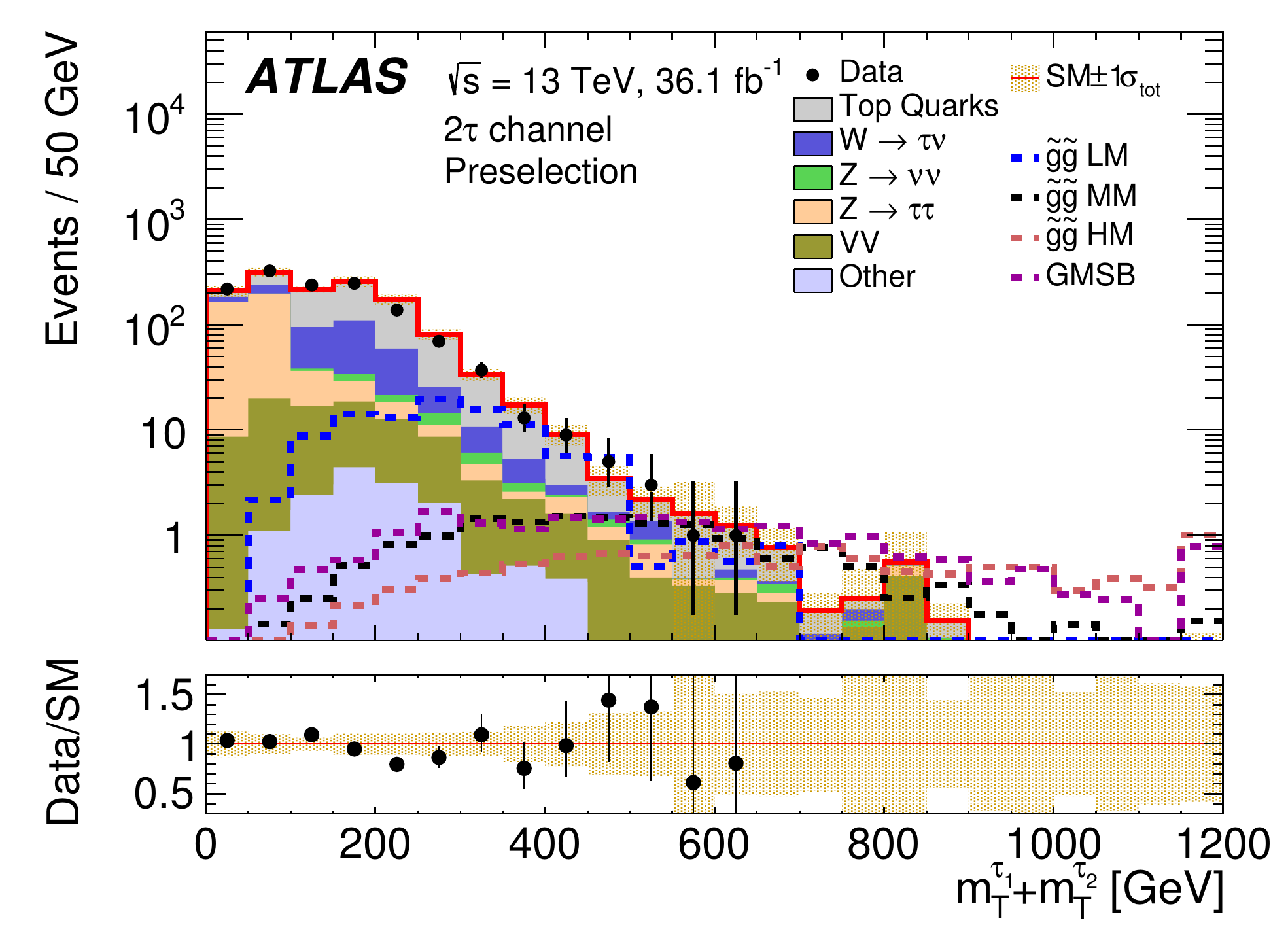}
                 } 
  \caption{Distributions of (a) the $\tau$-lepton transverse mass \mTtau in the \onetau channel and (b) the
sum of $\tau$-lepton transverse masses $m_\mathrm{T}^{\tau_1}+m_\mathrm{T}^{\tau_2}$ in the \twotau channel after the preselection, 
after applying data-driven normalization factors to the main backgrounds.
  The last bin includes overflow events. 
  The total uncertainty in the background prediction is shown as a shaded band.
  The contribution labeled as ``Other'' includes multijet events and the $V$+jets processes not explicitly listed in the legend.
Signal predictions are overlaid for several benchmark model points. 
For the simplified model, LM, MM, and HM refer to low, medium, and high mass-splitting scenarios, with $(m_{\tilde g},m_{\tilde\chi_1^0})$ set to $(1065,825)$~\gev, $(1625,905)$~\gev\ and $(1705,345)$~\gev, respectively. 
The GMSB benchmark model corresponds to $\Lambda=\SI{120}{\tera\electronvolt}$ and $\tan\beta=40$.  }
   \label{fig:preselection}
\end{figure}

Multiple phase space regions are then defined.
A set of signal regions (SRs) with stringent kinematic requirements and low background contribution is designed to target the different signatures and kinematic configurations of the two SUSY models.
A set of control regions (CRs) with negligible signal yield is used to constrain the normalization of the dominant backgrounds in phase space regions close to the SRs. 
The determination of background normalization factors and the search for a possible signal are performed simultaneously by fitting a signal-plus-background model to the data in the CRs and SRs.
Validation regions (VRs) are defined in phase space regions between CRs and SRs.
The VRs are not included in the fit; they are used to compare the fitted background predictions with the observed data in the vicinity of SRs to validate the background extrapolation before unblinding the SRs. 
The CRs, VRs and SRs are mutually exclusive and therefore statistically independent.

In the \onetau channel, two SRs are defined for the simplified model, as summarized in Table~\ref{tab:SR_1tau}.
The \onetau compressed SR targets small mass differences between the gluino and the LSP, up to $\approx\SI{300}{\giga\electronvolt}$. 
It exploits topologies where the pair of gluinos recoils against a high-\pt jet from initial-state radiation (ISR). 
While $\tau$-leptons and additional jets from gluino decays typically have low \pt, such ISR events have substantial \met since both LSPs tend to be emitted opposite to the ISR jet in the transverse plane. 
A requirement on the transverse mass is used to suppress $W(\tau\nu)$+jets events as well as semileptonic $t\bar t$ events with a $\tau$-lepton in the final state.
The \onetau medium-mass SR targets larger mass-splittings, motivating a more stringent $\mTtau$ criterion and an \HT requirement.

These two SRs also provide sensitivity to GMSB signals at low $\tan\beta$, in cases where only one $\tau$-lepton decays hadronically and is reconstructed within the detector acceptance.
 At high $\tan\beta$, the \onetau channel is not competitive due to the large multiplicity of $\tau$-leptons in signal events.

\begin{table}[htb]
  \caption{Summary of the signal region definitions in the \onetau channel.
  These requirements are applied in addition to the preselection. The variables are defined in the text.}
  \centering
  \begin{tabular}{l|c|c}
    \toprule
    Subject of & \multicolumn{2}{c}{\onetau SRs}\\
    selection & Compressed & Medium-mass\\
    \midrule
    $\tau$-leptons & $20 < \tauPt < \SI{45}{\giga\electronvolt}$ & $\tauPt > \SI{45}{\giga\electronvolt}$\\
    \midrule
    Event & \multicolumn{2}{c}{$\met > \SI{400}{\giga\electronvolt}$}\\
    kinematics & $\mTtau > \SI{80}{\giga\electronvolt}$ & $\mTtau > \SI{250}{\giga\electronvolt}$\\
    & --- & $\HT > \SI{1000}{\giga\electronvolt}$\\
    \bottomrule
  \end{tabular}
\label{tab:SR_1tau}
\end{table}

\pagebreak 

\begin{table}[htb]
  \caption{Summary of the signal region definitions in the \twotau channel.
  These requirements are applied in addition to the preselection. The variables are defined in the text.}
\small
  \centering
  \begin{tabular}{l|c|c|c|c}
    \toprule
    Subject of & \multicolumn{4}{c}{\twotau SRs}\\
    selection & Compressed & High-mass & Multibin & GMSB\\
    \midrule
    Event & $\mttwo^{\tau\tau} > \SI{70}{\giga\electronvolt}$ & $\mTTTkerned > \SI{350}{\giga\electronvolt}$ & $\mTTTkerned > \SI{150}{\giga\electronvolt}$ & $\mTTTkerned > \SI{150}{\giga\electronvolt}$\\
    kinematics & $\HT < \SI{1100}{\giga\electronvolt}$ & $\HT > \SI{1100}{\giga\electronvolt}$ & $\HT > \SI{800}{\giga\electronvolt}$ & $\HT > \SI{1900}{\giga\electronvolt}$\\
    & $\mT^\mathrm{sum} > \SI{1600}{\giga\electronvolt}$ & ---  & $\njet \geq 3$ & ---\\
    & & & 7 bins in \mTTTkerned & \\
    \bottomrule
  \end{tabular}
  \label{tab:SR_2tau}
\end{table}

In the \twotau channel, three SRs are defined for the simplified model, as summarized in Table~\ref{tab:SR_2tau}.
The compressed and high-mass SRs target signals with small and large mass-splittings, respectively.
The \twotau multibin SR exploits the shape difference between signal and background distributions, in contrast to the other SRs which only exploit the total yields. 
The multibin approach is less model-dependent than a single-bin SR designed to probe a narrow part of the model parameter space, and it provides increased sensitivity to both small and large mass-splittings.

The \twotau compressed SR has a requirement on $\mttwo^{\tau\tau}$ to exploit the kinematic endpoint of $Z(\tau\tau)$+jets and dileptonic $\ttbar$ events.
A requirement on $\mT^\mathrm{sum} \equiv \mTTT+ \sum\limits_i m_{\mathrm T}^{\mathrm {jet}_i}$ is imposed to take advantage of the large \met and the high multiplicities of jets and $\tau$-leptons that are expected from gluino decays and the boosted topologies. 
The upper bound on \HT allows a combination with the high-mass SR, and does not affect the sensitivity to compressed signals.
The \twotau high-mass SR includes a stringent requirement on $\mTTT$ that reduces the contribution from $Z(\tau\tau)$+jets events. The $\tau$-leptons from high-\pt $Z$ bosons have a small separation in $\phi$, which results in low values of \mTTT given that the $\tau$-neutrinos producing \met are collimated with the visible decay products of $\tau$-leptons.
An $\HT$ requirement is applied to significantly reduce background from $t\bar t$ and $W(\tau\nu)$+jets events.
The multibin SR uses looser selection criteria than the high-mass SR, and comprises seven bins in $\mTTT$.

A dedicated SR is defined for the GMSB model, based on the high-mass SR. 
To accommodate the more complex production and decay processes and the higher mass reach in the GMSB model, the minimum \mTTT requirement, which depends on specific decay topologies, is lowered while the minimum \HT requirement is raised.
The selection criteria defining the GMSB SR in the \twotau channel are summarized in Table~\ref{tab:SR_2tau}.

For the simplified model, the two SRs of the \onetau channel can be statistically combined in a simultaneous fit with either the compressed and high-mass SRs of the \twotau channel or the multibin SR of the \twotau channel, as the multibin SR is not mutually exclusive to the other \twotau SRs.
For each benchmark point in the parameter space, the most sensitive expected result of these two fits is used. 
For the GMSB interpretation, the \onetau SRs are combined with the \twotau GMSB SR and the \twotau compressed SR.

\section{Background estimation}
\label{sec:background}

Events from $W(\tau\nu)$+jets, $\ttbar$ and, to a smaller extent, diboson production are significant backgrounds in all SRs.
Additionally, $Z(\nu\nu)$+jets plays a role in the \onetau channel, while $Z(\tau\tau)$+jets is an important background in some of the \twotau SRs. Multijet production makes a minor contribution in the \onetau channel.
Dedicated control regions are used to constrain the normalization of all these backgrounds, except for diboson processes, which are normalized to their respective theoretical cross sections.

The $\tau$-leptons selected in the Standard Model background events are either prompt leptons from electroweak boson decays (\textit{true $\tau$-leptons}), or reconstructed objects such as jets that are misidentified as $\tau$-leptons (\textit{fake $\tau$-leptons}).
Backgrounds that contribute almost exclusively to a single channel, with only fake or only true $\tau$-leptons, are each normalized with a single normalization factor.
This is the case for $Z(\nu\nu)$+jets, multijet and $Z(\tau\tau)$+jets events.
The associated control regions are named $Z(\nu\nu)$ CR, multijet CR and $Z(\tau\tau)$ CR.
For both the $W(\tau\nu)$+jets and $\ttbar$ backgrounds, which contribute to both the \onetau and the \twotau SRs with different multiplicities of true and fake $\tau$-leptons, three normalization factors are used.
A normalization factor for true $\tau$-leptons is used to correct for differences in the $\tau$-lepton reconstruction and identification efficiencies between data and simulation.
A normalization factor for fake $\tau$-leptons accounts for multiple sources of potential mismodeling in the simulation: the quark/gluon composition of jets misidentified as $\tau$-leptons, the parton shower and hadronization models of the generator, and the modeling of particle shower shapes in the calorimeter, which mainly depends on the {\sc Geant4} hadronic interaction model and the modeling of the ATLAS detector.
An overall normalization factor accounts for the modeling of the background kinematics and acceptance, and absorbs the theoretical uncertainties in the cross-section computation, as well as the experimental uncertainties in the measured integrated luminosity of the data.
The corresponding CRs are named $W$/top true-$\tau$ CR, $W/$top fake-$\tau$ CR, and $W$/top kinematic CR, respectively.
The separation between $W$ and top CRs is achieved by requiring the absence or presence of a $b$-tagged jet.

The kinematic CRs require a muon and no $\tau$ candidate, to be independent of the $\tau$-lepton reconstruction and identification.
An upper bound on $m_\mathrm{T}^\mu$ is applied to select $W(\mu\nu)$+jets events and top-quark background with a muon in the final state.
The true-$\tau$ CRs target $W(\tau\nu)$+jets and semileptonic top-quark processes with a true $\tau$-lepton.
They are based on events with a $\tau$-lepton, jets, and \met.
Contributions from fake $\tau$-leptons are suppressed by a requirement on $m_\mathrm{T}^\tau$.
The fake-$\tau$ CRs target $W(\mu\nu)$+jets and top-quark processes with a final-state muon, with a jet misidentified as a $\tau$-lepton.
They use the same baseline selection as kinematic CRs, but a $\tau$ candidate is required. 
Events with large $m_\mathrm{T}^\mu$ values are discarded to suppress the top-quark background with a muon and a true $\tau$-lepton. In the $W$ fake-$\tau$ CR, the invariant mass of the reconstructed $\tau$-lepton and the muon  $m_{\tau\mu}$ is required to be large to suppress $Z(\tau\tau)$ events where one of the $\tau$-leptons decays into a muon.
The $Z(\nu\nu)$ CR requires one $\tau$-lepton, has a lower bound on $\mT^\tau$ to suppress background with real $\tau$-leptons, a requirement on $\met/\meff$, where $\meff=\HT+\met$, to reject multijet events, and requirements on the $\Delta\phi$ separations between the missing transverse momentum and the highest-\pt jet and $\tau$-lepton, to exploit the background topology.
The $Z(\tau\tau)$ CR is designed by inverting the \mTTT and \HT requirements from the \twotau SRs. 
This selection requires two medium $\tau$-leptons of opposite electric charge and imposes an upper bound on the invariant mass of the $\tau$-lepton pair to suppress dileptonic top-quark contributions.
Both $Z$ CRs employ a veto on $b$-tagged jets to suppress contributions from top-quark processes.
A simultaneous fit over all CRs is performed using {\sc HistFitter}~\cite{Baak:2014wma} to extract the normalization factors.

\begin{table}[tbp]
\caption{Summary of the $W$ and top control regions. 
These requirements are applied in addition to the trigger, jet, and multijet requirements of the preselection.
The variables $\ntau$, $\njet$, $N_{\mu}$ and $\nbjet$ are the number of $\tau$-leptons, jets, muons, and $b$-tagged jet, respectively; other variables are defined in the text.}
 \centering
\begin{tabular}{l|c|c|c}
\toprule
Subject of & $W$/~Top & $W$/~Top & $W$/~Top \\
selection & kinematic CR & true-$\tau$ CR & fake-$\tau$ CR\\
\midrule
$\tau$-leptons & $\ntau = 0$ & \multicolumn{2}{c}{$\ntau = 1$}\\
\midrule
Jets & \multicolumn{2}{c|}{$\njet \geq 3$} & ---\\
\midrule
Muons & $N_{\mu} = 1$ & $N_{\mu} = 0$ & $N_{\mu} = 1$\\
\midrule
$W$/top separation& \multicolumn{3}{c}{$\nbjet = 0/\geq 1$}\\
\midrule
Event & \multicolumn{3}{c}{$\HT <  800 \gev$}\\
kinematics & \multicolumn{3}{c}{$\met < 300 \gev$}\\ 
& $\mTmu <  100 \gev$ & $\mTtau < 80 \gev$ & $\mTmu <  100 \gev$ \\
& --- & --- & $m_{\tau\mu} >  60 \gev$ ($W$ CR) \\
\bottomrule
\end{tabular}
\label{tab:bkg:CRsone}
\end{table}

\begin{table}[tbp]
\caption{Summary of the $Z(\nu\nu)$, $Z(\tau\tau)$ and multijet control regions.
These requirements are applied in addition to the trigger, and jet requirements of the preselection.
The variables $\ntau$ and $N_{\mu}$ are the number of $\tau$-leptons, and muons, respectively;
 $q_{\tau_i}$ is the charge of $\tau$-lepton $i$; other variables are defined in the text.}
 \centering
\begin{tabular}{l|c|c|c}
\toprule
Subject of selection & $Z(\nu\nu)$ CR & $Z(\tau\tau)$ CR & Multijet CR \\
\midrule
$\tau$-leptons & $\ntau = 1$ & $\ntau \ge 2,\,q_{\tau_1}=-q_{\tau_2}$ & $\ntau = 1$\\
\midrule
Multijet events & \multicolumn{2}{c|}{$\dphijets > 0.4$} & $\dphijets < 0.3$\\
\midrule
Muons &  $N_{\mu} = 0$ & --- & ---\\
\midrule
Top suppression &  \multicolumn{2}{c|}{$\nbjet = 0$} & ---\\
\midrule
Event kinematics & \multicolumn{2}{c|}{$\HT <  800 \gev$} & ---\\
& $\met < 300 \gev$ & --- & ---\\ 
&  $ 100 \leq \mTtau <  200 \gev $ & $\mTTT <  100 \gev$ & $100 < \mTtau < 200 \gev$ \\
&  $\metovermeff > 0.3$ & $\mttwo <  70 \gev$ & $\metovermeff <0.2$ \\
&  $\dphiLjet > 2.0$ & ---& ---\\
&  $\dphiLtau > 1.0$ & --- & ---\\
\bottomrule
\end{tabular}
\label{tab:bkg:CRstwo}
\end{table}

The multijet background contributes when jets are misidentified as $\tau$-leptons and large missing transverse momentum is induced by jet energy mismeasurements.
This, together with the very large production cross section, makes it difficult to simulate a sufficient number of multijet events with the required accuracy, so this background is estimated from data~\cite{SUSY-2011-20}.
A data sample with high purity in multijet events is selected using single-jet triggers.
Events with well-measured jets are retained by applying an upper bound on the \met significance~\cite{SUSY-2016-01}, except for events where the leading $b$-tagged jet is aligned with \ptmissv. The latter exception avoids too large of a suppression of high-\pt $b$-hadrons decaying semileptonically and producing high-\pt neutrinos.
Jet energies are then smeared according to the jet energy resolution obtained from simulation and corrected to better describe the data.
The smearing is performed multiple times for each selected event, leading to a large pseudo-data set where \met originates from resolution effects and which includes an adequate fraction of jets misidentified as $\tau$-leptons. 
A subtraction is performed to account for the small contamination from $\ttbar$ events satisfying this kinematic configuration.
The normalization of the pseudo-data is constrained in the simultaneous fit using a multijet CR where either of the two leading jets is aligned with $\boldsymbol{p}_{\mathrm{T}}^{\mathrm{miss}}$.

\begin{figure}[p]
\centering
\subfigure[]{\includegraphics[width=0.50\textwidth]{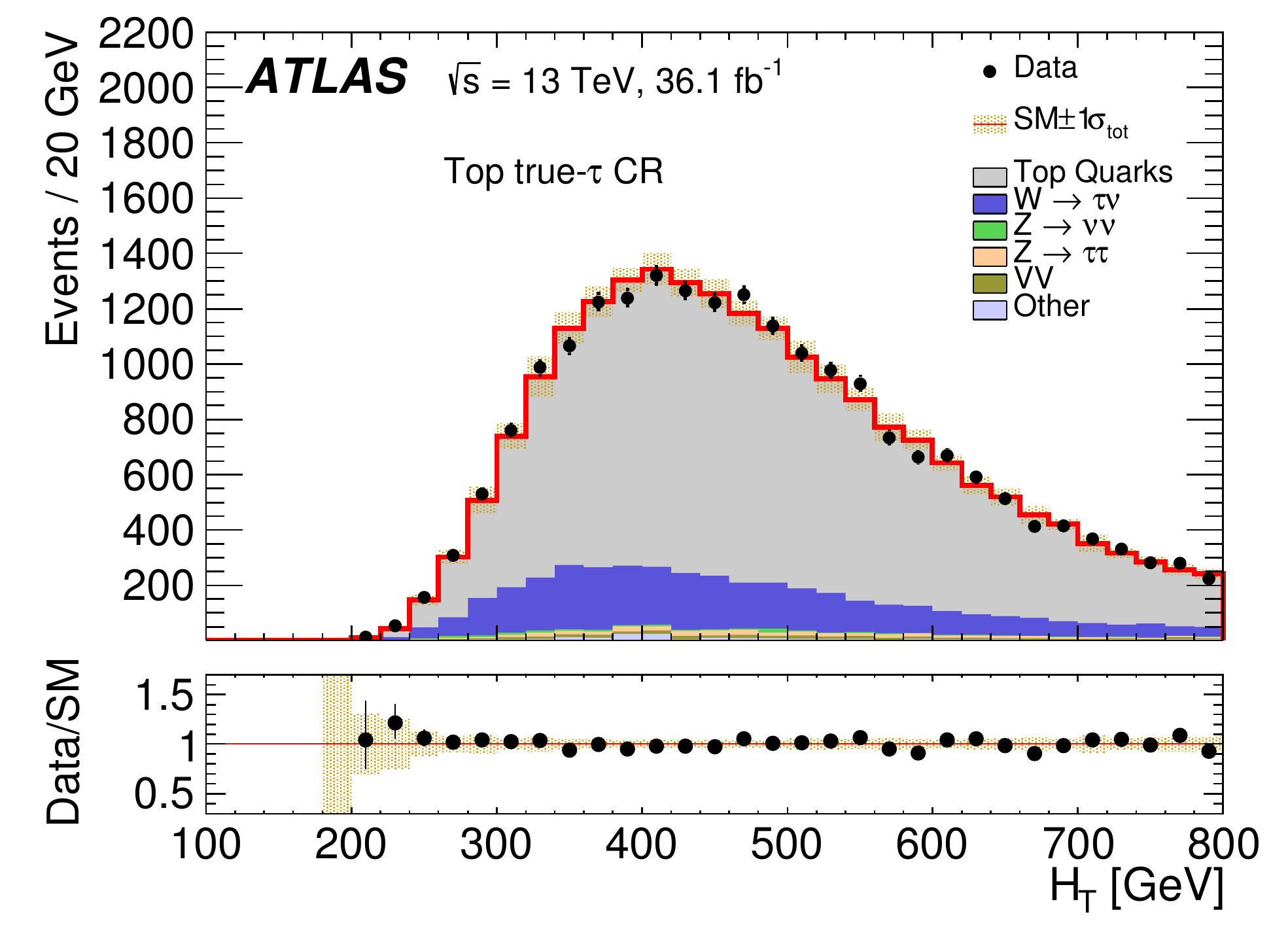}}
\hspace*{-1ex}\subfigure[]{\includegraphics[width=0.50\textwidth]{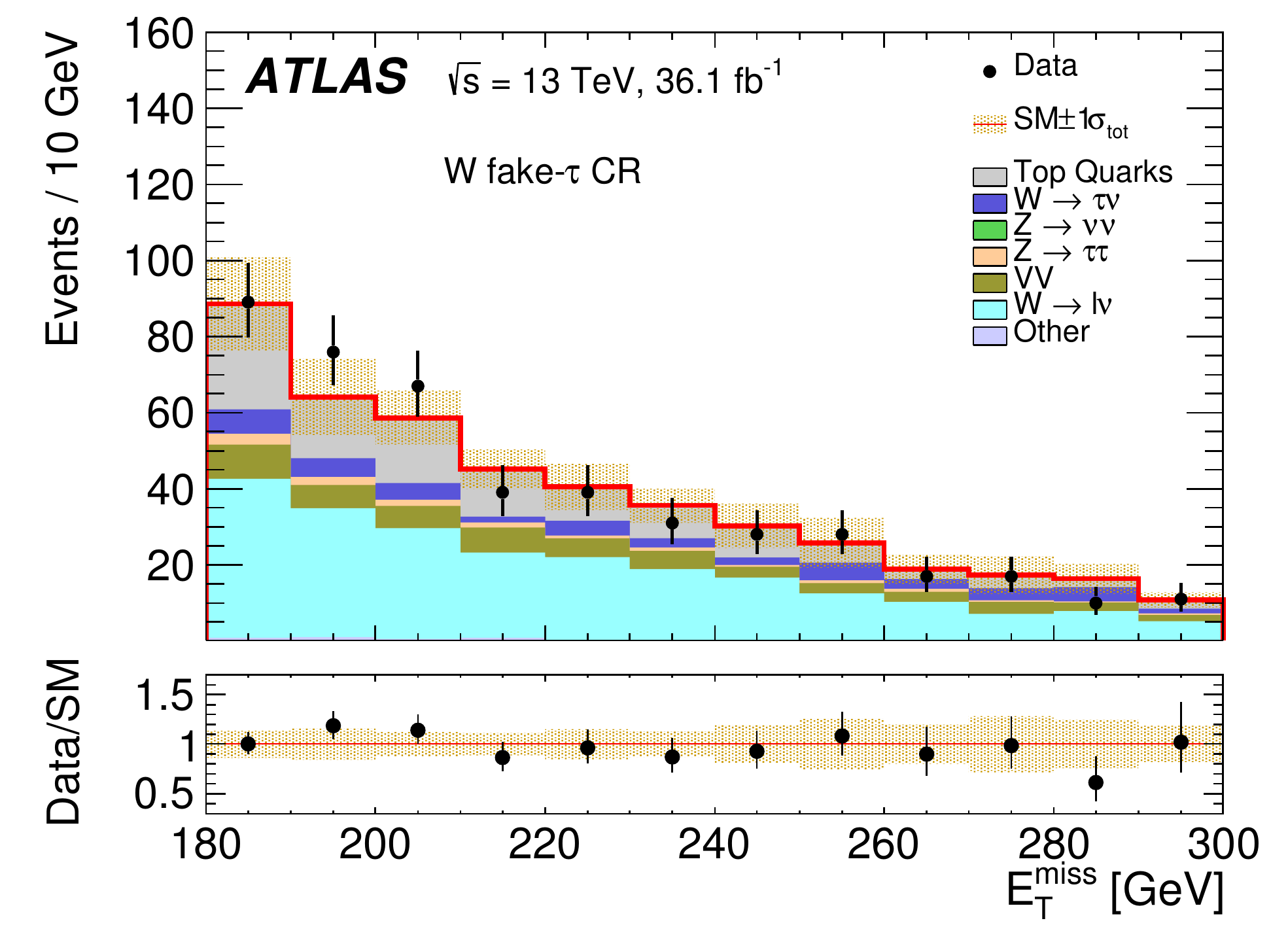}}
\hspace*{-1ex}\subfigure[]{\includegraphics[width=0.50\textwidth]{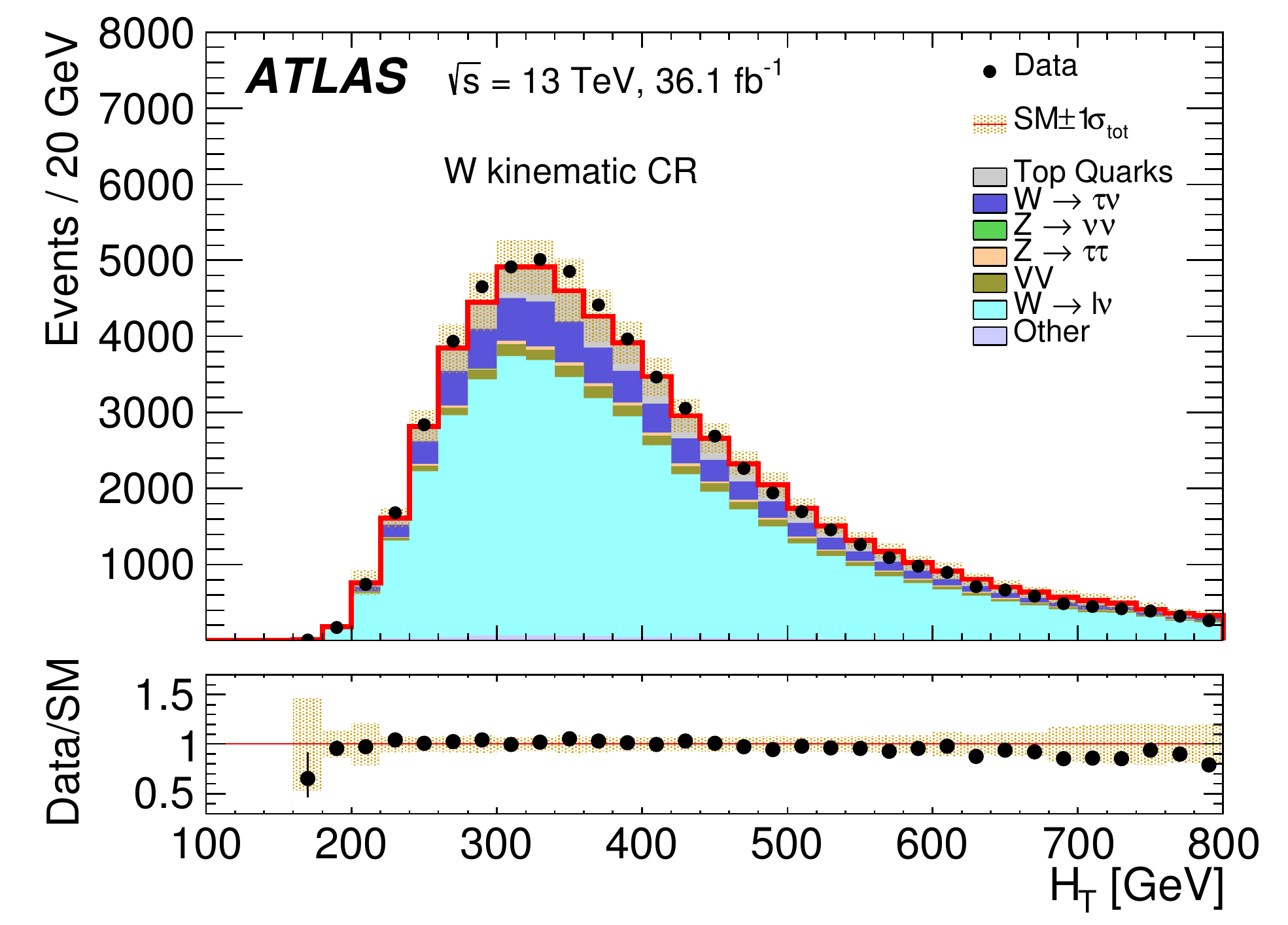}}
\hspace*{-1ex}\subfigure[]{\includegraphics[width=0.50\textwidth]{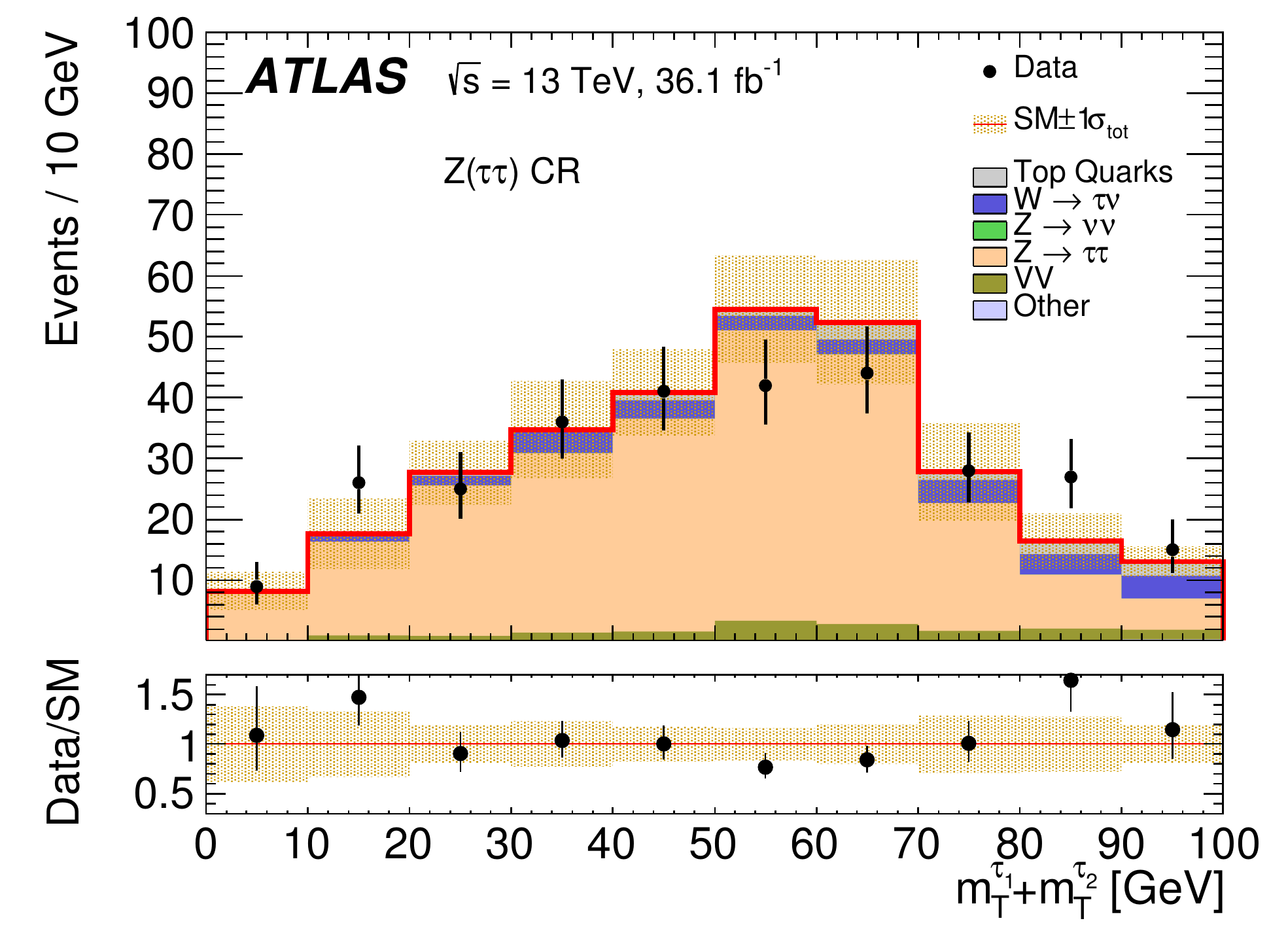}}
\hspace*{-1ex}\subfigure[]{\includegraphics[width=0.50\textwidth]{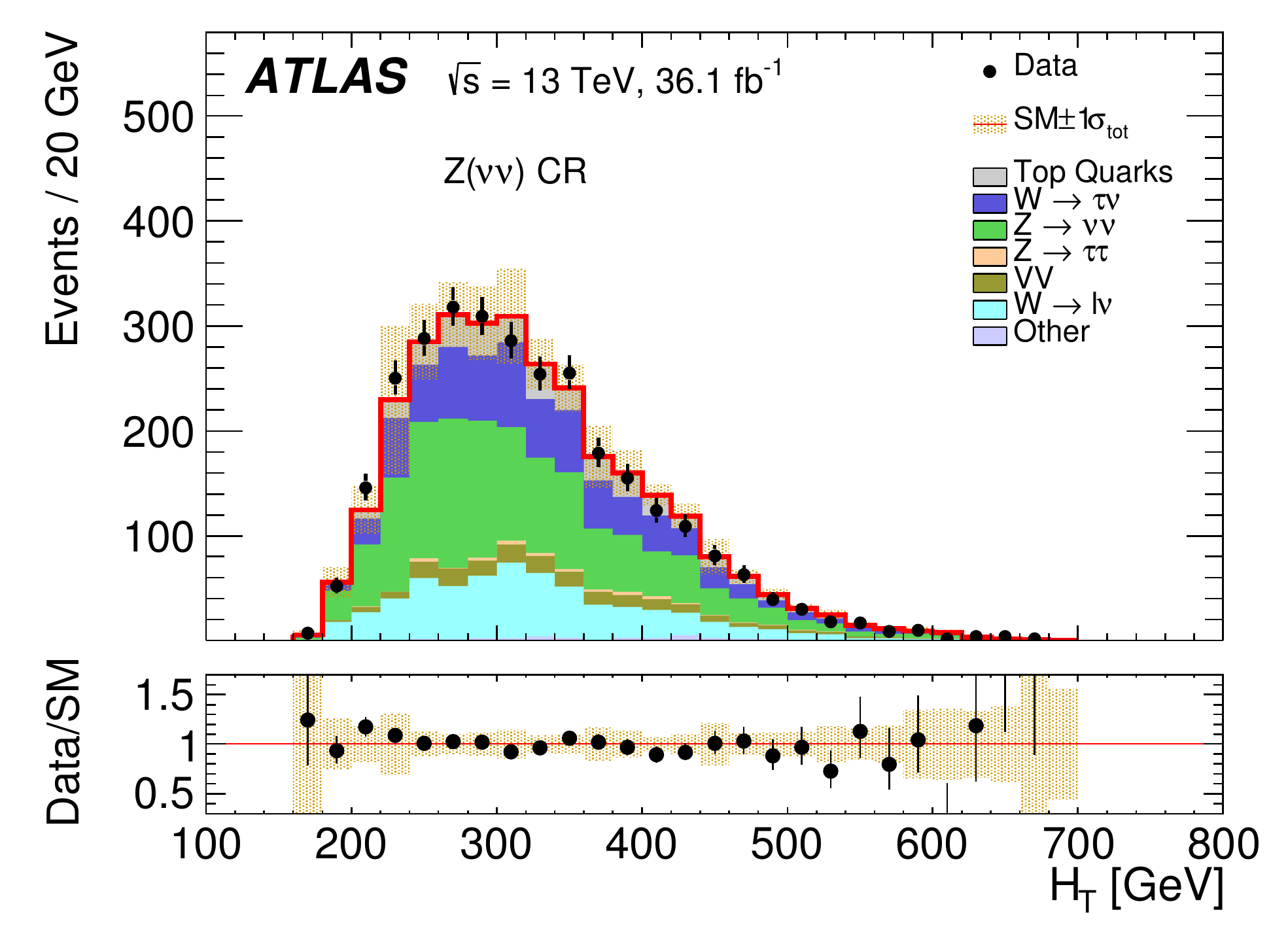}}
\hspace*{-1ex}\subfigure[]{\includegraphics[width=0.50\textwidth]{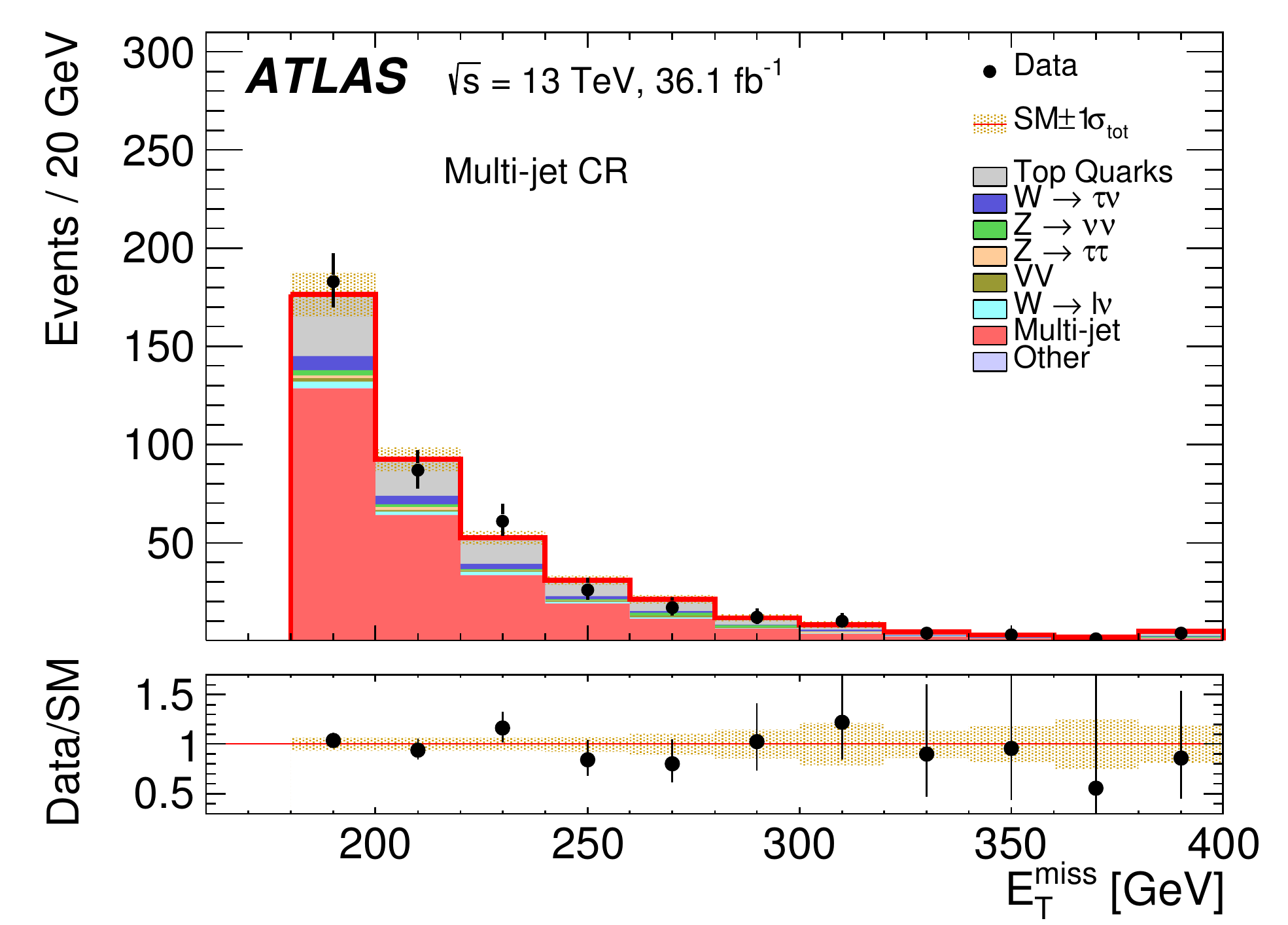}}
\caption{(a) Scalar sum of transverse momenta of $\tau$-leptons and jets \HT in the top true-$\tau$ CR, (b) missing transverse momentum \met in the $W$ fake-$\tau$ CR, (c) \HT in the $W$ kinematic CR, (d) sum of $\tau$-lepton transverse masses \mTTTkerned in the $Z(\tau\tau)$ CR, (e) \HT in the $Z(\nu\nu)$ CR, and (f) \met in the multijet CR, illustrating the background modeling in the CRs after the fit. The contribution labeled as ``Other'' includes multijet events (except for the multijet CR) and the $V$+jets processes not explicitly listed in the legend.
The last bin of each distribution includes overflow events. 
The total uncertainty in the background prediction is shown as a shaded band.
}
\label{fig:bkg:CRplots}
\end{figure}

The selection criteria defining the various CRs are summarized in Tables~\ref{tab:bkg:CRsone} and \ref{tab:bkg:CRstwo}.
Figure~\ref{fig:bkg:CRplots} illustrates the background modeling in CRs after the fit.
The fitted normalization factors do not deviate from unity by more than 15\% and are compatible with unity within one standard deviation when considering all systematic uncertainties, except for the $Z(\nu\nu)$+jets background, where the normalization factor reaches $1.44\pm0.29$.

Validation regions are used to verify that the background is well modeled after the fit in kinematic regions close to the SRs.
In the \onetau channel, three VRs are defined for the medium-mass SR and two for the compressed SR, while three VRs are used for the \twotau channel.
Their selection criteria are summarized in Tables~\ref{tab:ew_onetau:VRs} and \ref{tab:ew_twotau:VRs}.
The level of agreement between data and background in the VRs is illustrated in Figures~\ref{fig:bkg:VRplots_1T} and \ref{fig:bkg:VRplots_2T}.
Distributions are found to be well modeled in both channels.
The comparison between the numbers of observed events and the predicted background yields is displayed in Figure~\ref{fig:bkg:VRpulls}.
Agreement well within one standard deviation is observed.

\begin{table}[b]
  \caption{Validation regions for the \onetau channel.
These requirements are applied in addition to the preselection.
The variables are defined in the text.}
  \centering
  \begin{tabular}{l|c|c|c|c|c}
    \toprule
Subject of & \multicolumn{3}{c|}{\onetau medium-mass VRs} & \multicolumn{2}{c}{\onetau compressed VRs}\\
selection & $\HT$ & \met & $\mTtau$ & \met & $\mTtau$\\
    \midrule
    $\tau$-leptons & \multicolumn{3}{c|}{$\tauPt > \SI{45}{\giga\electronvolt}$} & \multicolumn{2}{c}{$20 < \tauPt < \SI{45}{\giga\electronvolt}$}\\
    \midrule
    Event& \multicolumn{2}{c|}{$\mTtau < \SI{250}{\giga\electronvolt}$} & $\mTtau > \SI{250}{\giga\electronvolt}$ & $\mTtau < \SI{80}{\giga\electronvolt}$ & $\mTtau > \SI{80}{\giga\electronvolt}$\\
    kinematics & $\met < \SI{400}{\giga\electronvolt}$ & $\met > \SI{400}{\giga\electronvolt}$ & $\met < \SI{400}{\giga\electronvolt}$ & $\met > \SI{400}{\giga\electronvolt}$ & $\met < \SI{400}{\giga\electronvolt}$\\
    & $\HT > \SI{1000}{\giga\electronvolt}$ & \multicolumn{2}{c|}{$\HT < \SI{1000}{\giga\electronvolt}$} & --- & ---\\	
    \bottomrule
  \end{tabular}
  \label{tab:ew_onetau:VRs}
\end{table}

\begin{table}[b]
  \caption{Validation regions for the \twotau channel.
These requirements are applied in addition to the preselection.
The variables are defined in the text.}
  \centering
  \begin{tabular}{l|c|c}
    \toprule
Subject of selection & \twotau $W$/Top VR& $Z(\tau\tau)$ VR\\
    \midrule
    $W$/top separation & $\nbjet = 0/\geq 1$ & --- \\
    \midrule
    Event & $\HT < \SI{800}{\giga\electronvolt}$ & $\HT > \SI{800}{\giga\electronvolt}$\\
    kinematics & $\mTTT > \SI{150}{\giga\electronvolt}$ & $\mTTT < \SI{150}{\giga\electronvolt}$\\
    & \multicolumn{2}{c}{$\mttwo^{\tau\tau} < \SI{60}{\giga\electronvolt}$} \\
    \bottomrule
  \end{tabular}
  \label{tab:ew_twotau:VRs}
\end{table}

\begin{figure}
\centering
\subfigure[]{\includegraphics[width=0.50\textwidth]{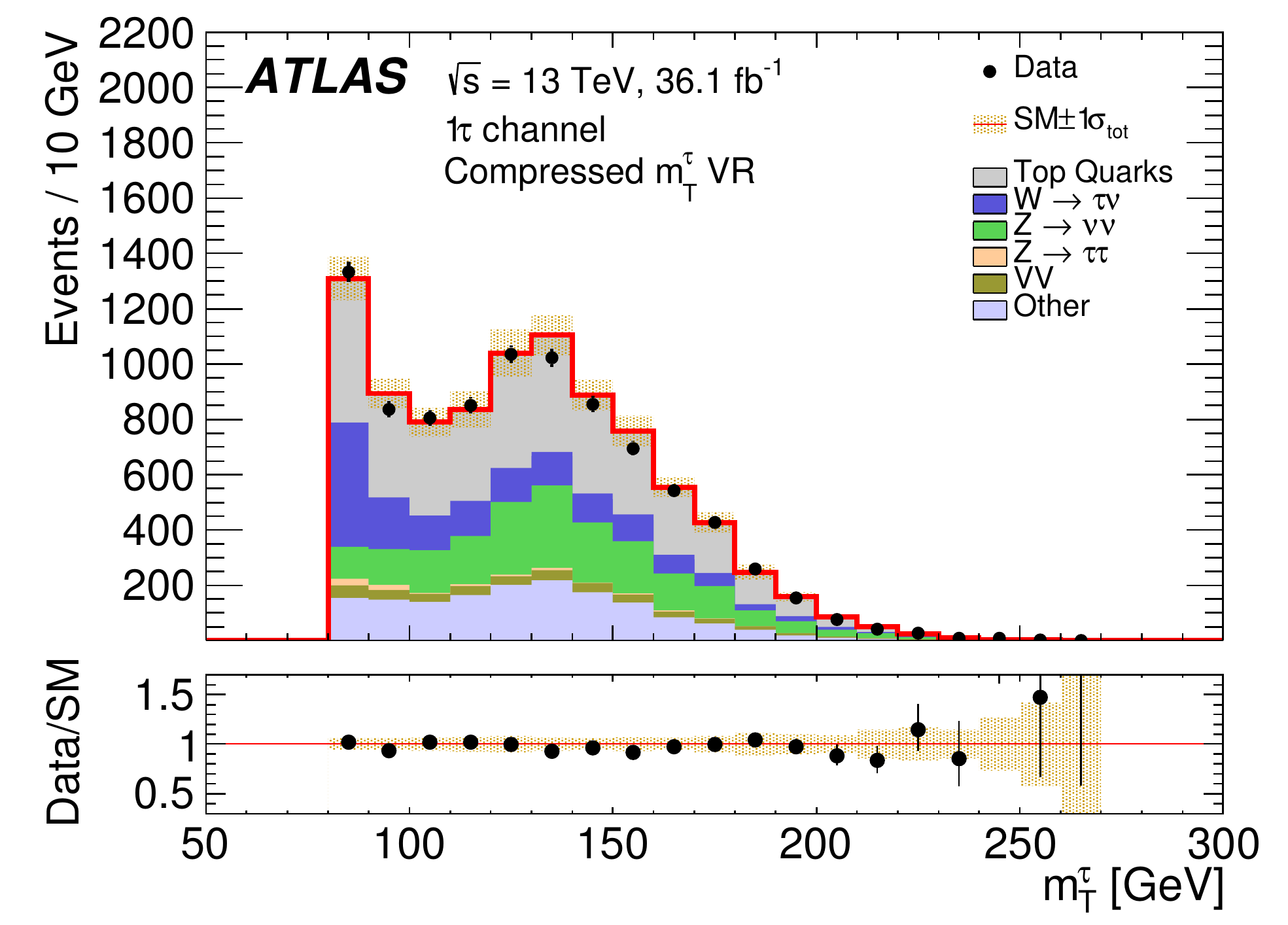}}
\hspace*{-1ex}\subfigure[]{\includegraphics[width=0.50\textwidth]{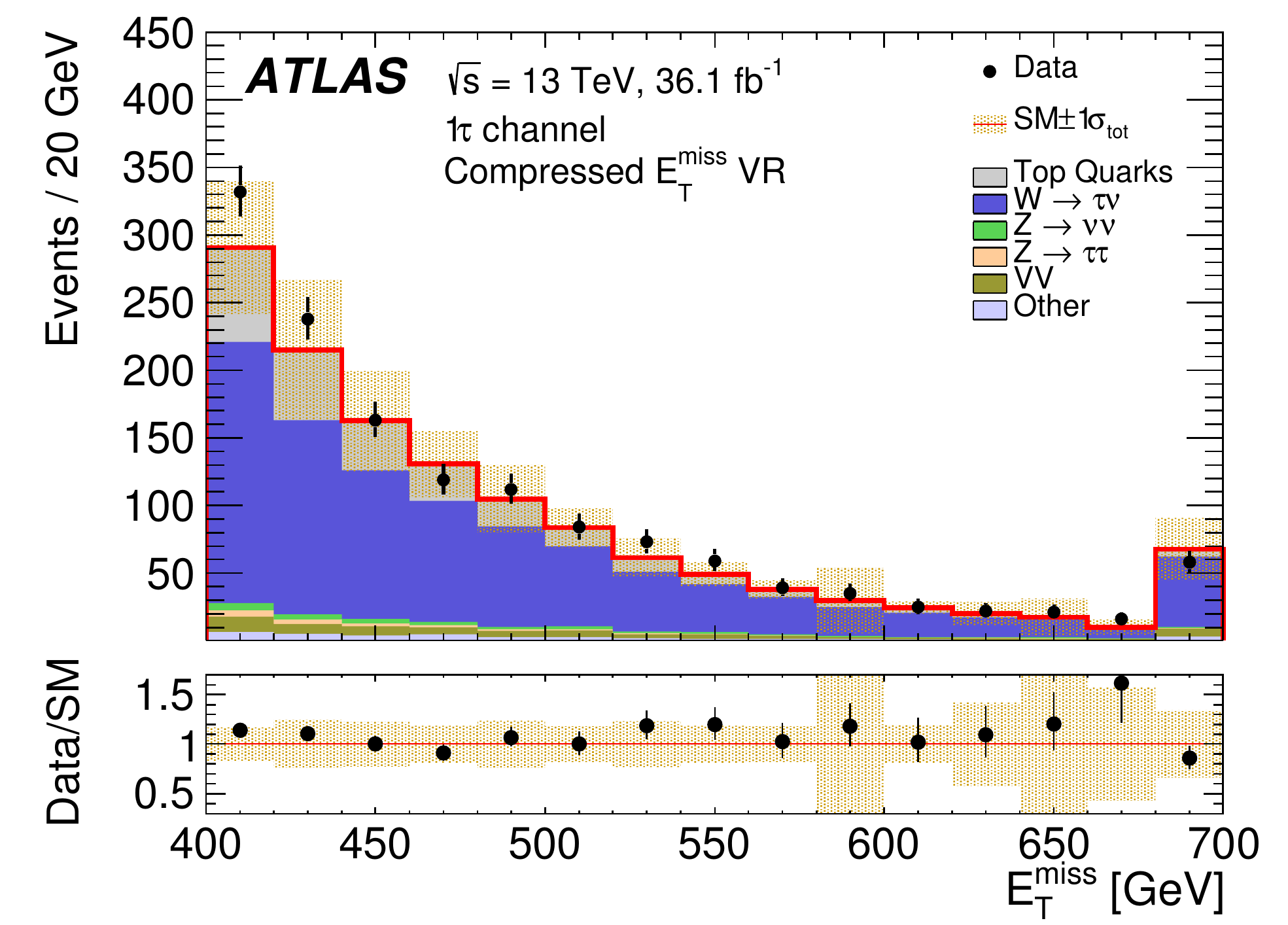}}
\hspace*{-1ex}\subfigure[]{\includegraphics[width=0.50\textwidth]{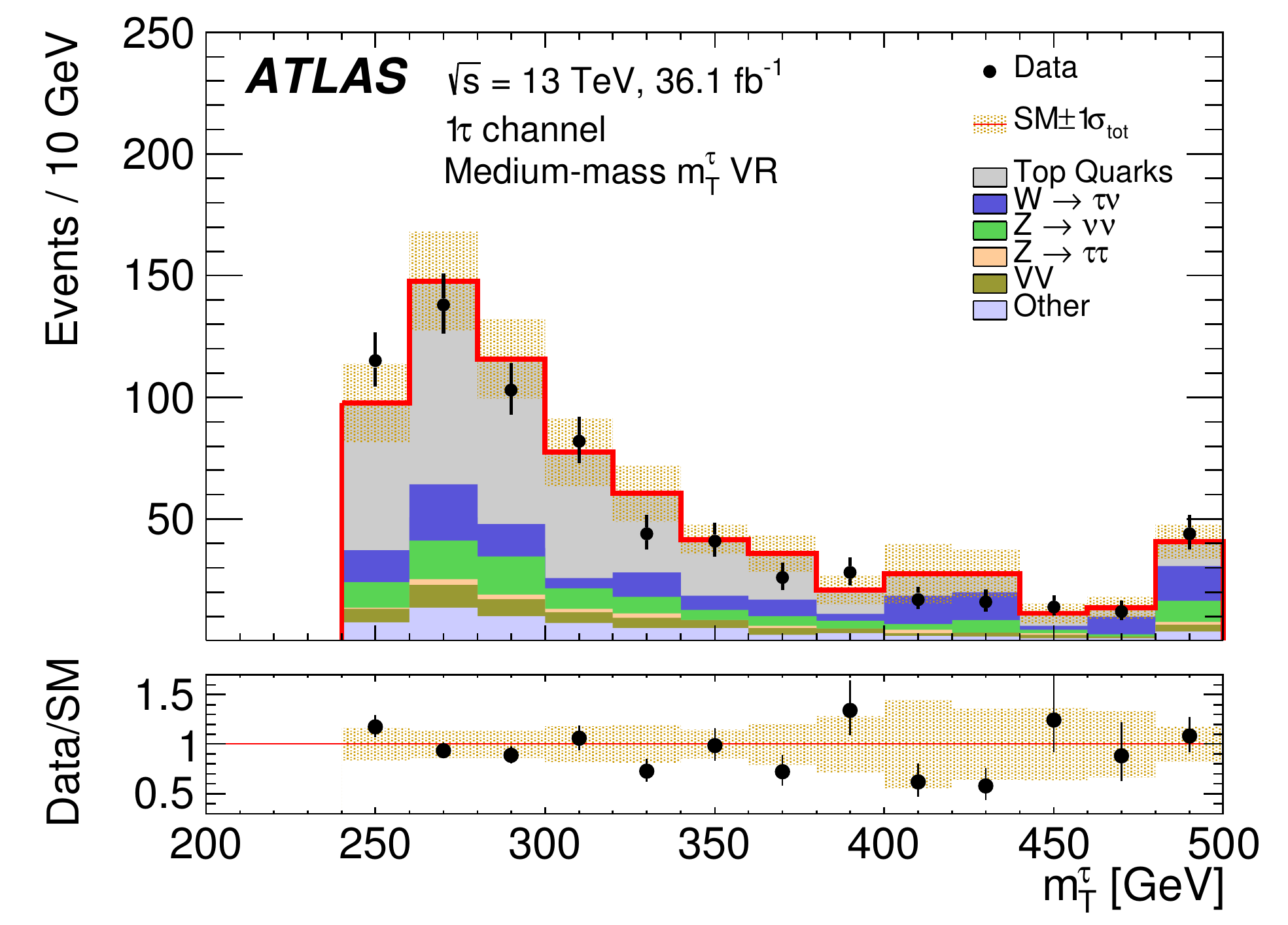}}
\hspace*{-1ex}\subfigure[]{\includegraphics[width=0.50\textwidth]{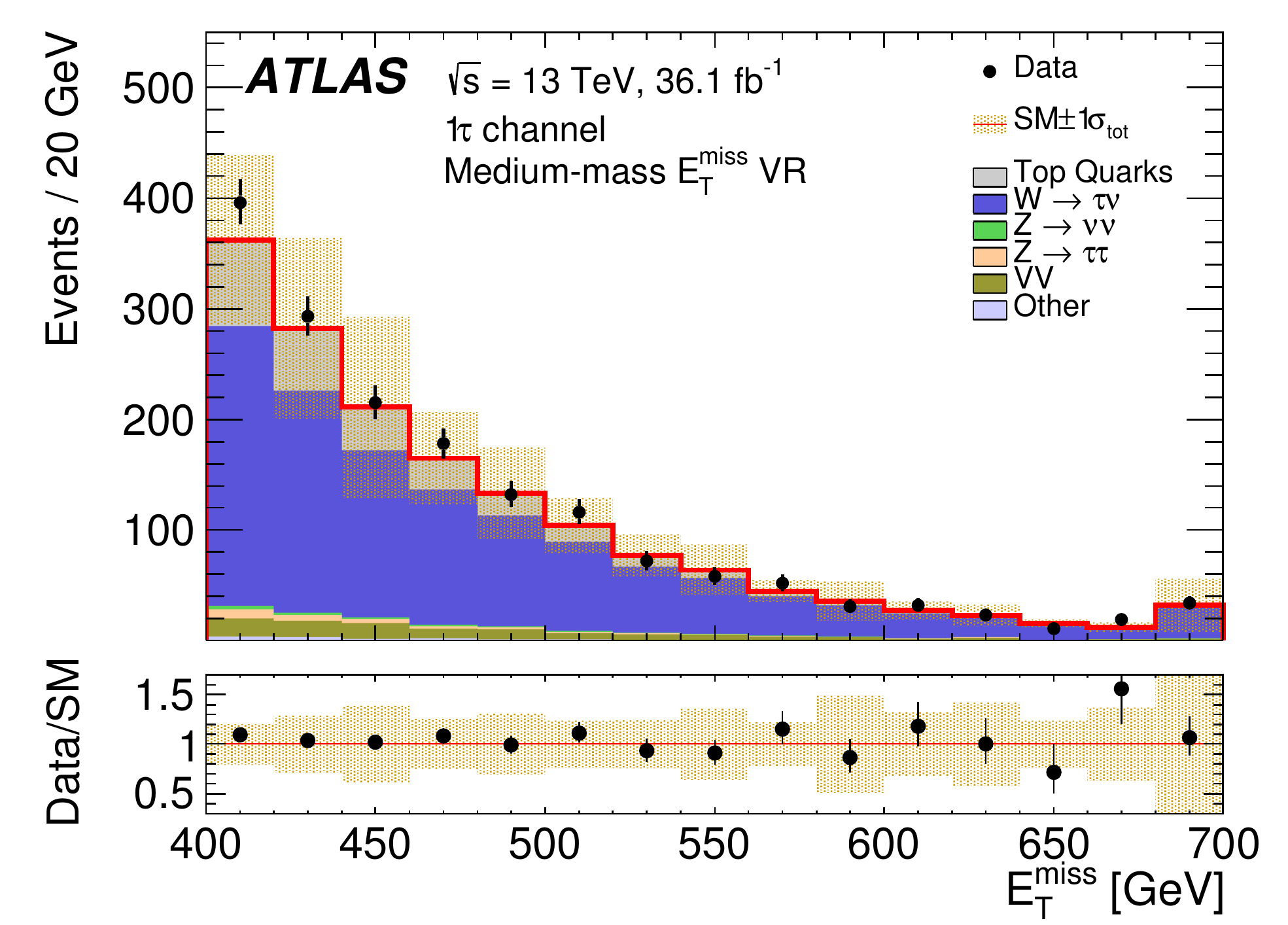}}
\hspace*{-1ex}\subfigure[]{\includegraphics[width=0.50\textwidth]{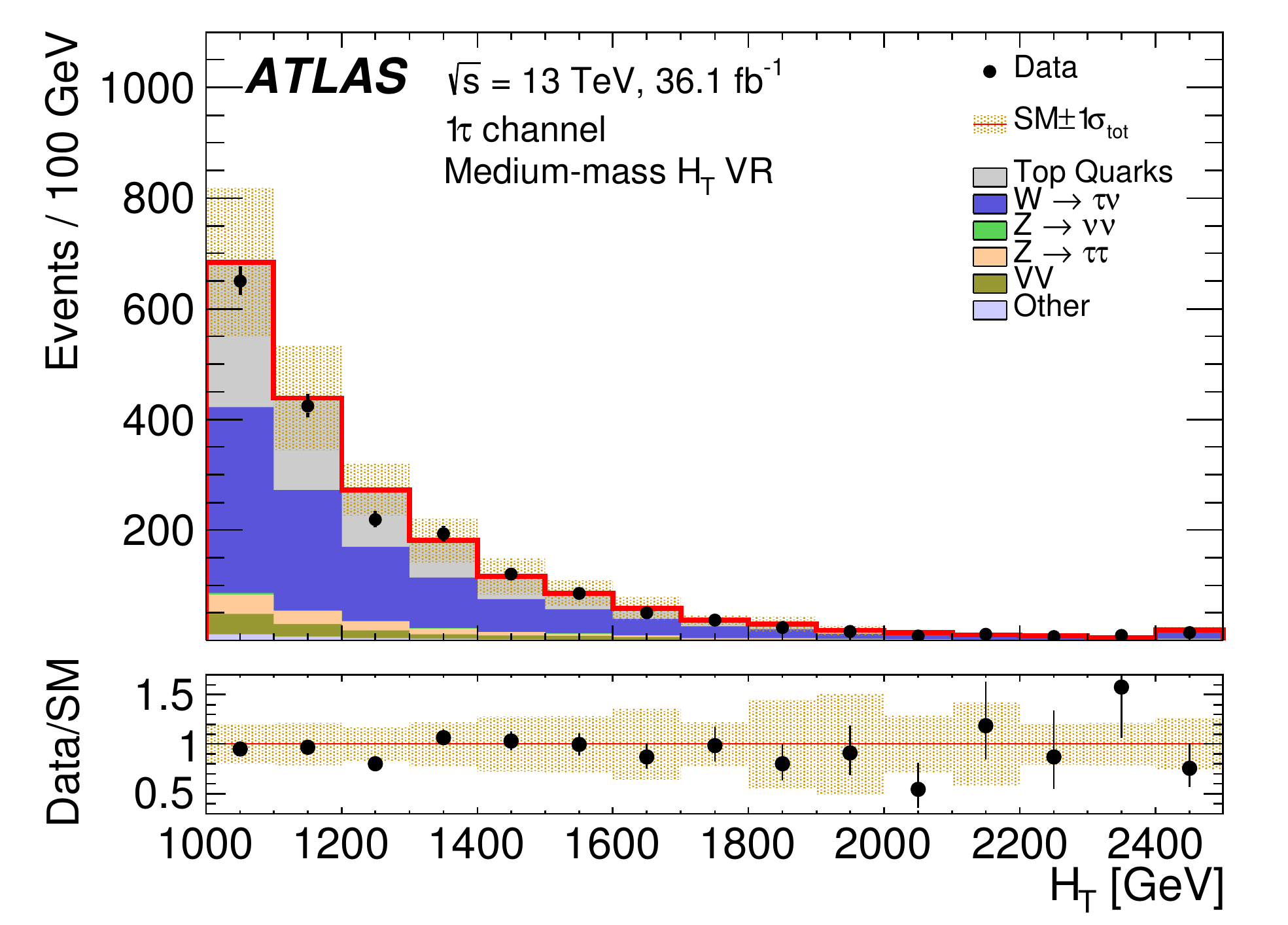}}
\caption{Distributions of (a) $\tau$-lepton transverse mass \mTtau in the compressed \mTtau VR, (b) missing transverse momentum \met in the compressed \met VR, (c) \mTtau in the medium-mass \mTtau VR, (d) \met in the medium-mass \met VR, and (e) scalar sum of $\tau$-lepton and jet transverse momenta \HT in the medium-mass \HT VR, 
 illustrating the background modeling in the VRs of the \onetau channel after the fit. The normalization factors obtained in the CRs are applied.
The contribution labeled as ``Other'' includes multijet events and the $V$+jets processes not explicitly listed in the legend.
The last bin of each distribution includes overflow events. 
The total uncertainty in the background prediction is shown as a shaded band.
}
\label{fig:bkg:VRplots_1T}
\end{figure}

\begin{figure}
\centering
\subfigure[]{\includegraphics[width=0.50\textwidth]{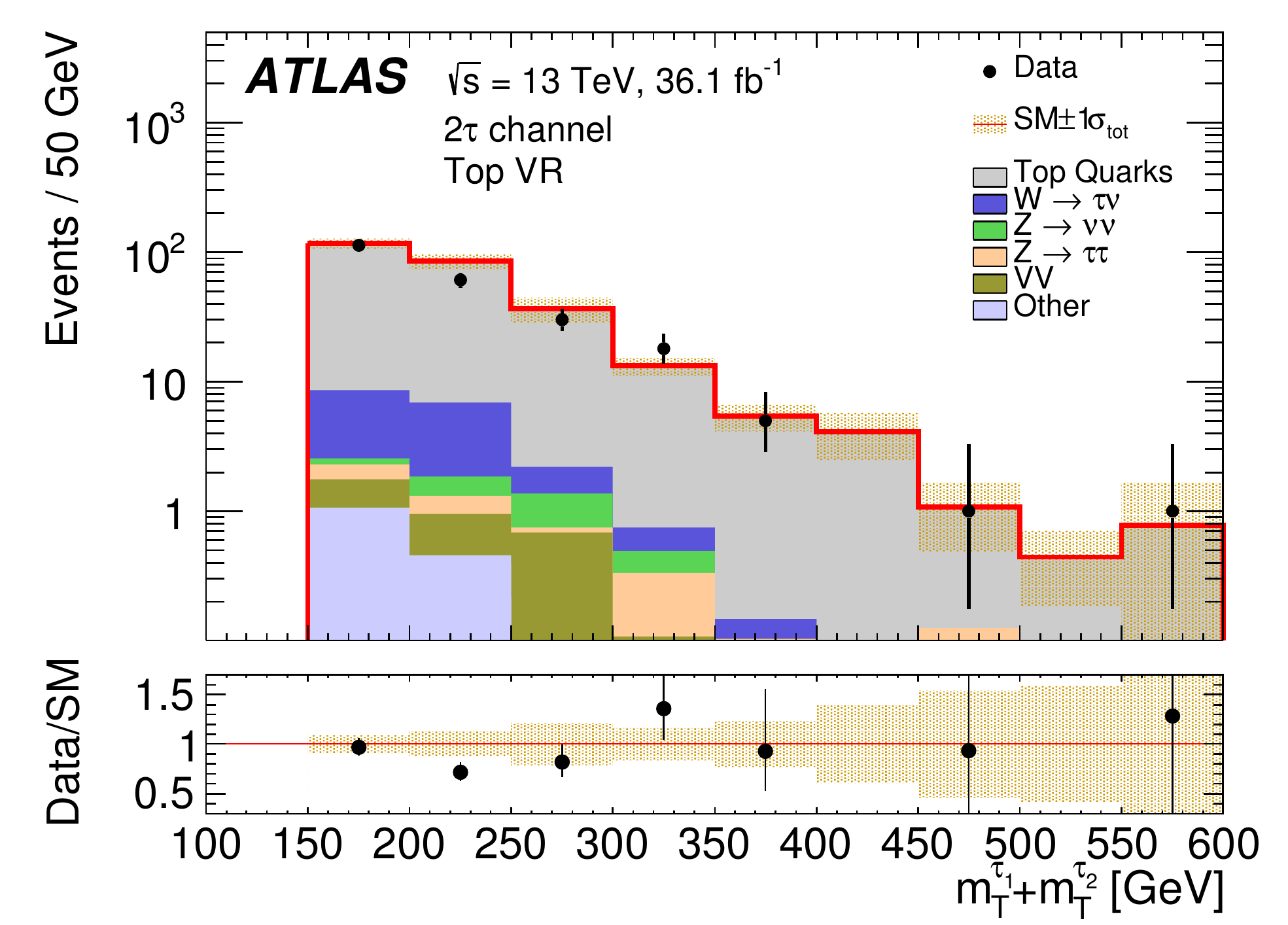}}
\hspace*{-1ex}\subfigure[]{\includegraphics[width=0.50\textwidth]{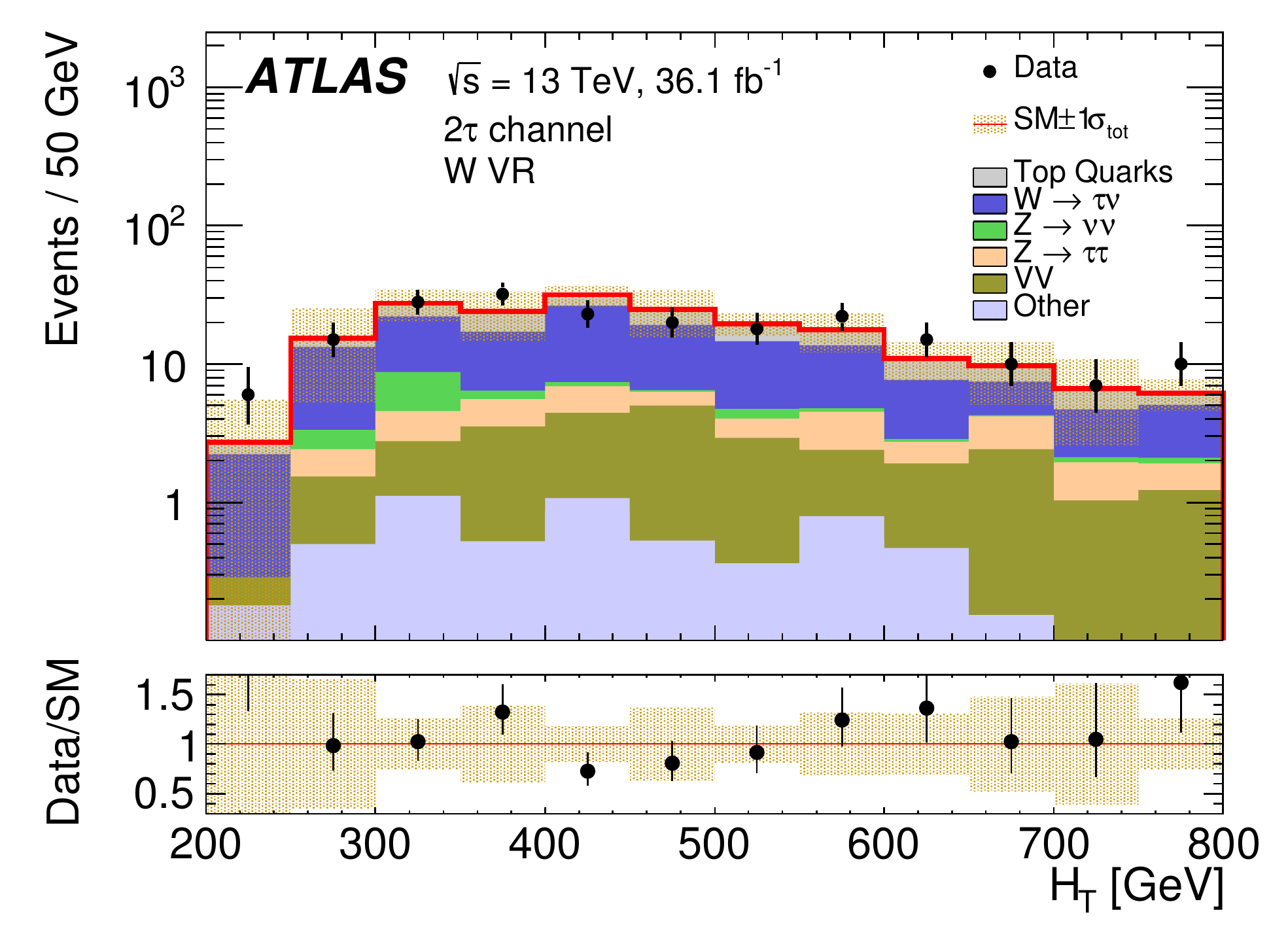}}
\hspace*{-1ex}\subfigure[]{\includegraphics[width=0.50\textwidth]{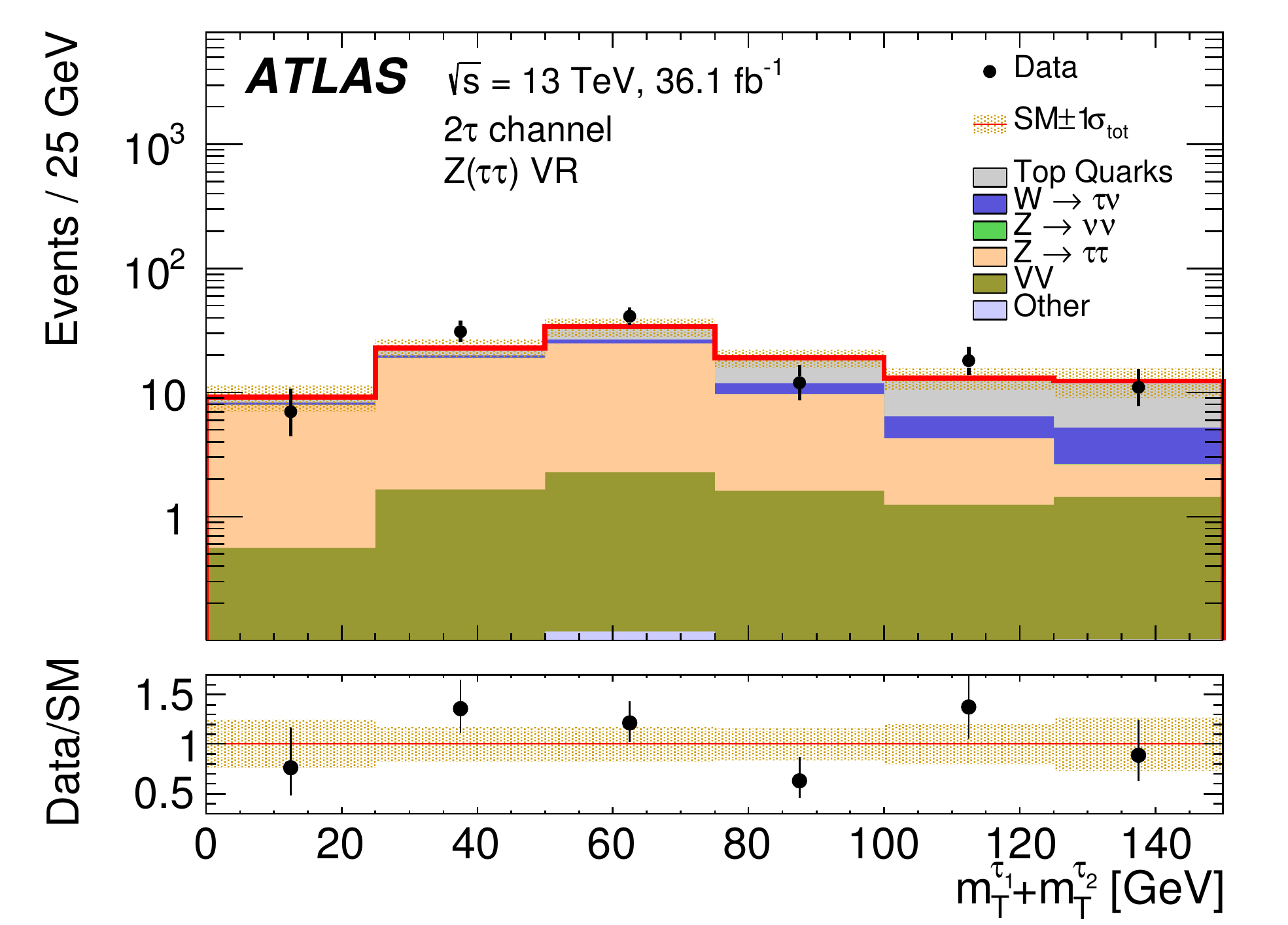}}
\caption{(a) Sum of $\tau$-lepton transverse masses \mTTTkerned in the top VR, (b) scalar sum of $\tau$-lepton and jet transverse momenta \HT in the $W$ VR, and (c) \mTTTkerned in the $Z$ VR,
illustrating the background modeling in the VRs of the \twotau channel after the fit. The normalization factors obtained in the CRs are applied. 
The contribution labeled as ``Other'' includes multijet events and the $V$+jets processes not explicitly listed in the legend.
The last bin of each distribution includes overflow events. 
The total uncertainty in the background prediction is shown as a shaded band.
}
\label{fig:bkg:VRplots_2T}
\end{figure}

\begin{figure}
\centering
\includegraphics[width=\textwidth]{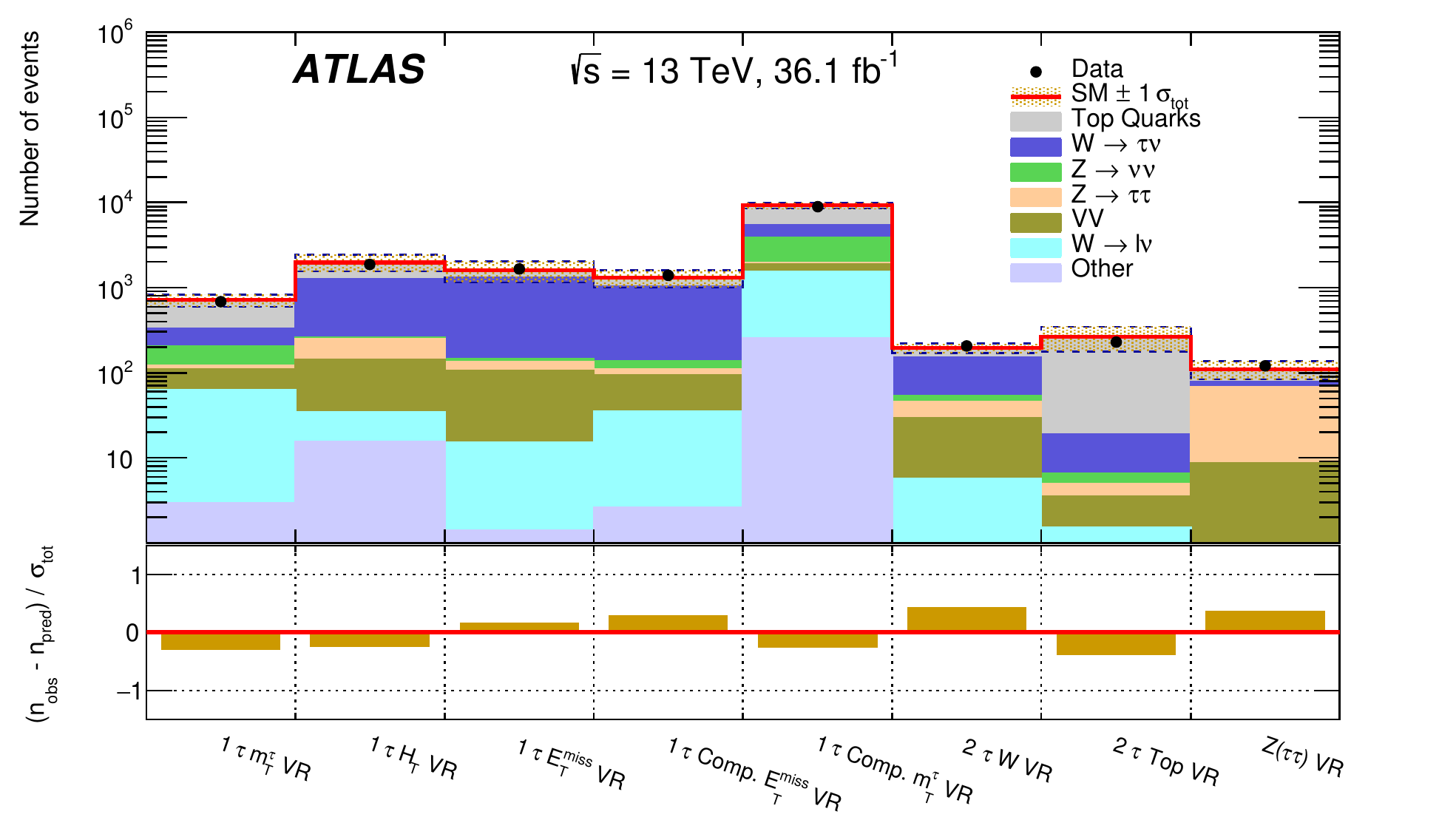}
\caption{Number of observed events $n_\mathrm{obs}$ and predicted background yields in the validation regions $n_\mathrm{pred}$ of the \onetau and \twotau channels.
The background predictions are scaled using normalization factors derived in the control regions. 
 The total uncertainty in the background predictions $\sigma_\mathrm{tot}$ is shown as a shaded band. The lower panel displays the significance of the deviation of the observed from the expected yield.
}
\label{fig:bkg:VRpulls}
\end{figure}

\FloatBarrier

\section{Systematic uncertainties}
\label{sec:systematics}

Theoretical and experimental systematic uncertainties are evaluated for all simulated processes.
The uncertainties from theory include PDF, $\alphas$ and scale uncertainties, and generator modeling uncertainties.
Experimental uncertainties are related to the reconstruction, identification, and calibration of final-state objects.
Specific uncertainties are evaluated for the multijet background, which is estimated from data.

For $V$+jets and diboson samples, systematic uncertainties related to PDFs, $\alphas$, and scales are evaluated using alternative weights from the generator.
The PDF uncertainty is obtained as the standard deviation of the 100 PDF variations from the NNPDF3.0nnlo set. 
The effect of the uncertainty in $\alphas$ is computed as half the difference resulting from the $\alphas=0.119$ and $\alphas=0.117$ parameterizations.
The renormalization scale $\mu_{\mathrm{R}}$ and factorization scale $\mu_{\mathrm{F}}$ are varied up and down by a factor of two and all combinations are evaluated, except for the $(2\mu_{\mathrm{R}},\frac{1}{2}\mu_{\mathrm{F}})$ and $(\frac{1}{2}\mu_{\mathrm{R}},2\mu_{\mathrm{F}})$ variations, which would lead to large $\log(\mu_{\mathrm{R}}/\mu_{\mathrm{F}})$ contributions to the cross section. 
The scale uncertainty is computed as half the difference between the two combinations yielding the largest and smallest deviations from the nominal prediction.
Uncertainties due to the resummation and CKKW matching scales for $V$+jets samples are found to be negligible. 
Additional generator modeling uncertainties are considered for the dominant $W(\tau\nu)$+jets background.
An uncertainty is derived to cover a mismodeling of the \HT distribution observed in the $W$ kinematic CR (cf. Figure~\ref{fig:bkg:CRplots}\,(c)).
In addition, predictions from {\sc Sherpa} and {\sc MG5\_aMC@NLO + Pythia8} are compared, and the difference is taken as a systematic uncertainty.
For the diboson background, which is not normalized to data in the fit, the uncertainty in the cross section is also taken into account.

For top quark pair production, uncertainties due to PDF and scale variations are derived using {\sc Powheg + Pythia8} and applied to the nominal predictions from {\sc Powheg + Pythia6}. 
Generator modeling uncertainties are assessed from comparisons with alternative generator samples.
An uncertainty in the hard-scattering model is evaluated by comparing predictions from {\sc MG5\_aMC@NLO} + \Herwigpp and {\sc Powheg-Box} + \Herwigpp.
An uncertainty due to the parton shower and hadronization models is evaluated by comparing predictions from {\sc Powheg-Box + Pythia6} and {\sc Powheg-Box} + \Herwigpp.
An uncertainty due to the ISR modeling is assessed by varying the {\sc Powheg-Box} parameter which controls the transverse momentum of the first additional parton emission beyond the Born configuration. 
For the small contributions from single-top-quark production and $\ttbar+V$ events, uncertainties in the cross sections are taken into account.

Systematic uncertainties affecting jets arise from the jet energy scale~\cite{PERF-2016-04}, jet energy resolution~\cite{PERF-2011-04}, and efficiency corrections for jet-vertex-tagging~\cite{PERF-2014-03} 
as well as $b$-tagging~\cite{ATL-PHYS-PUB-2016-012}.
Jet energy scale uncertainties are mainly determined from measurements of the \pT balance in the calorimeter in $Z/\gamma$+jet and multijet events.
Remaining uncertainties arise from the relative calibration of forward and central jets, jet flavor composition, pileup, and punch-through for high-\pt jets not fully contained in the calorimeters.
A set of five uncertainties that comprises contributions from both absolute and in situ energy calibrations and which preserves the dominant correlations in the $(\pt,\eta)$ phase space is used.
An uncertainty in the jet energy resolution is applied to jets in the simulation as a Gaussian energy smearing.

Systematic uncertainties affecting true $\tau$-leptons are related to the reconstruction and identification efficiencies, the electron rejection efficiency, and the energy scale calibration~\cite{ATL-CONF-2017-029}.
The uncertainties in the reconstruction efficiency are estimated by varying parameters in the simulation such as the detector material, underlying event, and hadronic shower model.
Uncertainties in the identification efficiency and in situ energy calibration, which are derived in $Z(\tau\tau)$ events with a hadronically decaying $\tau$-lepton and a muon, 
arise from the modeling of true- and fake-$\tau$-lepton templates.
The uncertainty in the energy scale also includes non-closure of the calibration found in simulation and a single-pion response uncertainty.
In the case of fake $\tau$-leptons, the misidentification rate in the simulation is largely constrained by the fit to data in the CRs.
The process-dependence of the misidentification rate is accounted for by the use of different normalization factors for the various backgrounds.
Uncertainties in the extrapolation from the CRs to the VRs and SRs are covered by generator modeling uncertainties.

In the case of signal samples, which undergo fast calorimeter simulation, dedicated uncertainties take into account the difference in performance between full and fast simulation.
These uncertainties include non-closure of the energy calibration for both the jets and $\tau$-leptons, as well as differences in reconstruction and identification efficiencies for $\tau$-leptons.

Systematic uncertainties in the missing transverse momentum originate from uncertainties in the energy or momentum calibration of jets, $\tau$-leptons, electrons, and muons, which are propagated to the \met calculation.
Additional uncertainties are related to the calculation of the track-based soft term. 
These uncertainties are derived by studying the \pt balance between the soft term and the hard term 
composed of all reconstructed objects, in $Z(\mu\mu)$ events. 
Soft-term uncertainties include scale uncertainties along the hard-term axis, and resolution uncertainties along and perpendicular to the hard-term axis~\cite{ATL-PHYS-PUB-2015-023}.

A systematic uncertainty accounts for the modeling of pileup in the simulation, which affects the correlation between the average number of interactions per bunch crossing and the number of reconstructed primary vertices.
The modeling mostly depends on the minimum-bias tune and the longitudinal size of the $pp$ interaction region used in the simulation. 

Systematic uncertainties in the small multijet background contribution are due to the limited numbers of events in the input data set satisfying the \met significance requirement, the jet resolution parameterization used for jet energy smearing, and the $\ttbar$ background subtraction.

The uncertainty in the combined 2015+2016 integrated luminosity is 2.1\%. It is derived, following a methodology similar to that detailed in Ref.~\cite{DAPR-2013-01}, from a calibration of the luminosity scale using $x$--$y$ beam-separation scans performed in August 2015 and May 2016.

The impact of the main systematic uncertainties on the total background predictions in the SRs of the \onetau and \twotau channels is summarized in Table~\ref{tab:systematics}.
These uncertainties are shown after the background fit, assuming that no signal is present in the CRs.
In both channels, generator modeling uncertainties for the $W$+jets and $\ttbar$ backgrounds are the largest sources of systematic uncertainty. 
Other dominant uncertainties are jet energy calibration and $\tau$-lepton identification, which contributes more in the \twotau channel.
Uncertainties in the $b$-tagging efficiency and \met calibration have little impact on background predictions, and those affecting electrons and muons are negligible.

\begin{table}[tb]
\caption{Dominant systematic uncertainties in the total background predictions, for the signal regions of the \onetau (top) and \twotau (bottom) channels after the normalization fit in the control regions. 
The total systematic uncertainty accounts for other minor contributions not listed in this table. Due to non-trivial correlations between the various sources in the combined fit, the total uncertainty is not identical to the sum in quadrature of the individual components.}
\begin{center}
\setlength{\tabcolsep}{0.0pc}
\begin{tabular*}{\textwidth}{@{\extracolsep{\fill}}lrrr}
\toprule
Source of uncertainty	  & \onetau compressed SR         & \onetau medium-mass SR          \\
\midrule
Top generator modeling &6\%& 11\% \\
$V$+jets generator modeling &7\%& 5\% \\
Jet energy scale and resolution &7\%& 7\% \\
$\tau$-lepton energy scale 	&$<1\%$& 2.9\% \\ 
$\tau$-lepton identification 	&1.5\%& 3.3\% \\ 
PDFs 	& 1.9\%& 13\%\\ 
Limited simulation sample size	& 1.8\%	& 6\%	\\
Background normalization uncertainty	& 12\%	& 11\%	\\
\midrule
Total & $ 10\% $ & $ 19\% $ \\
\bottomrule
\\
\toprule
Source of uncertainty	  & \twotau compressed SR         & \twotau high-mass SR            & \twotau GMSB SR             \\
\midrule
Top generator modeling &31\%& 18\%&14\% \\
$V$+jets generator modeling &7\%& 15\%&21\% \\
Jet energy scale and resolution &15\%& 9\%&5\% \\
$\tau$-lepton energy scale 	&4\%& 6\%&1.7\% \\ 
$\tau$-lepton identification 	&5\%& 10\%&9\% \\ 
PDFs 	& 2.0\%&4\%& 10\%\\ 
Limited simulation sample size	& 10\%	& 8\%	& 21\%\\
Background normalization uncertainty	& 13\%	& 13\%	& 13\%	\\
\midrule
Total 				& $35\% $	& $30\%$	& $38\%$	\\
\bottomrule
\end{tabular*}
\label{tab:systematics}
\end{center}
\end{table}

\section{Results}
\label{sec:result}

Kinematic distributions for the SRs of the \onetau and \twotau channels are shown in Figures~\ref{fig:onetau:n_1} and~\ref{fig:twotau:n_1}, respectively. In these plots, all selection criteria defining the respective SRs are applied, except for the one on the variable which is displayed. 
Data and fitted background predictions are compared, and signal predictions from several benchmark models are overlaid. Variables providing the most discrimination between signal and background are displayed. The \mTTTkerned distribution which is used for the multibin SR of the \twotau channel is also shown.

Good agreement between data and background expectation is observed. A small discrepancy is observed for $\mTtau < \SI{200}{\giga\electronvolt}$ in the \onetau compressed SR (cf. Figure~\ref{fig:onetau:n_1:mt}). This region has been studied in detail and no particular problem has been identified. Given that the deviation is only observed in a restricted region and it is below two standard deviations in all bins, no significant impact on the result is expected.

The numbers of observed events and expected background events in the SRs of the \onetau and \twotau channels are reported in Tables~\ref{tab:results_mutual:yields:SR1T} and \ref{tab:results_mutual:yields:SR2T}, respectively.
In the high-mass and GMSB SRs of the \twotau channel that both require high \HT, a small excess of data with a significance of below two standard deviations is observed. 
Apart from that, no significant deviation of data from the SM prediction is observed in any of the five single-bin SRs and the seven bins of the multibin SR. Upper limits are set at the 95\% confidence level (CL) on the number of signal events, or equivalently, on the signal cross section.

\begin{figure}
\centering
\subfigure[]{\includegraphics[width=0.50\textwidth]{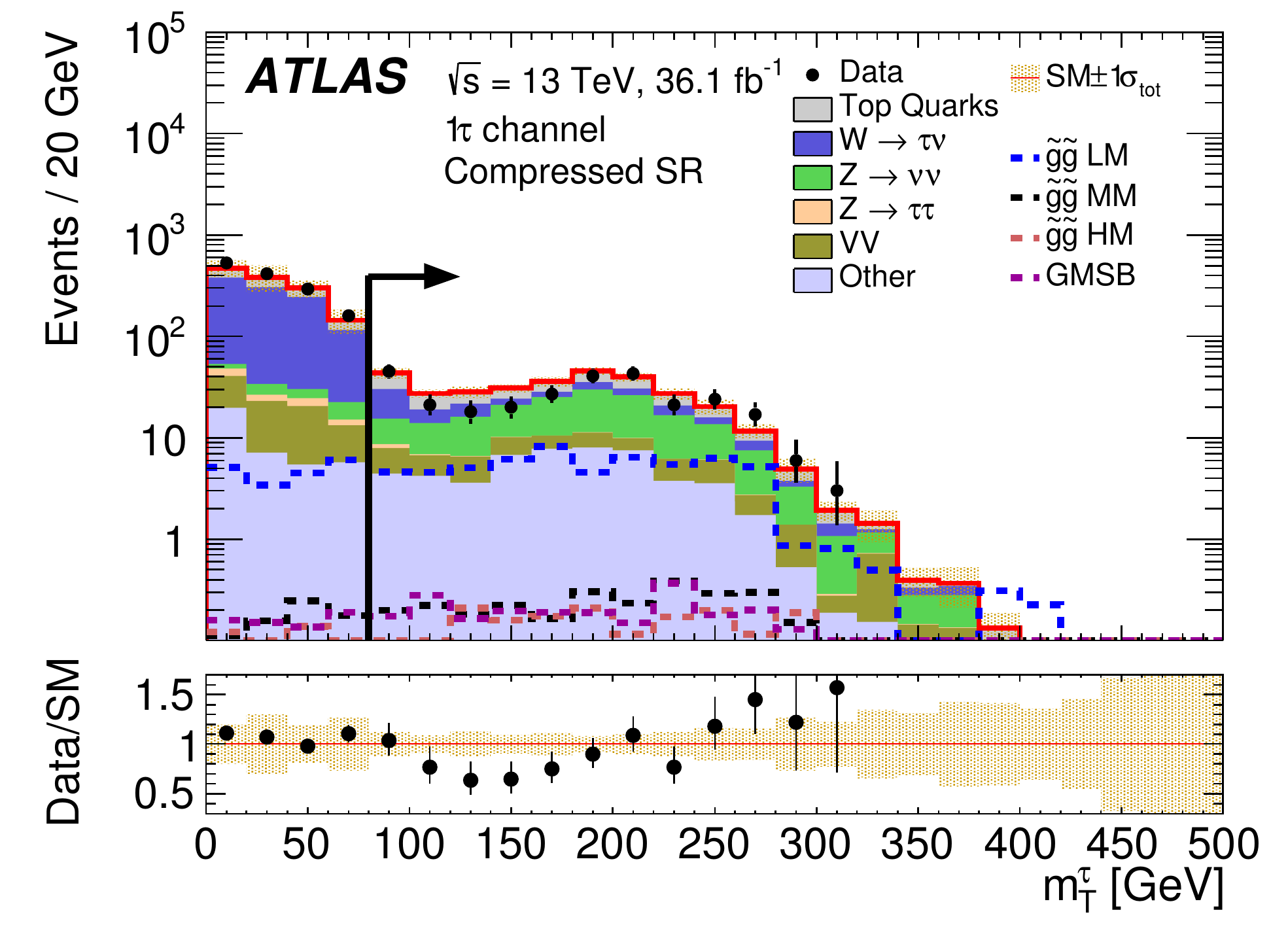}\label{fig:onetau:n_1:mt}}
\hspace*{-1ex}\subfigure[]{\includegraphics[width=0.50\textwidth]{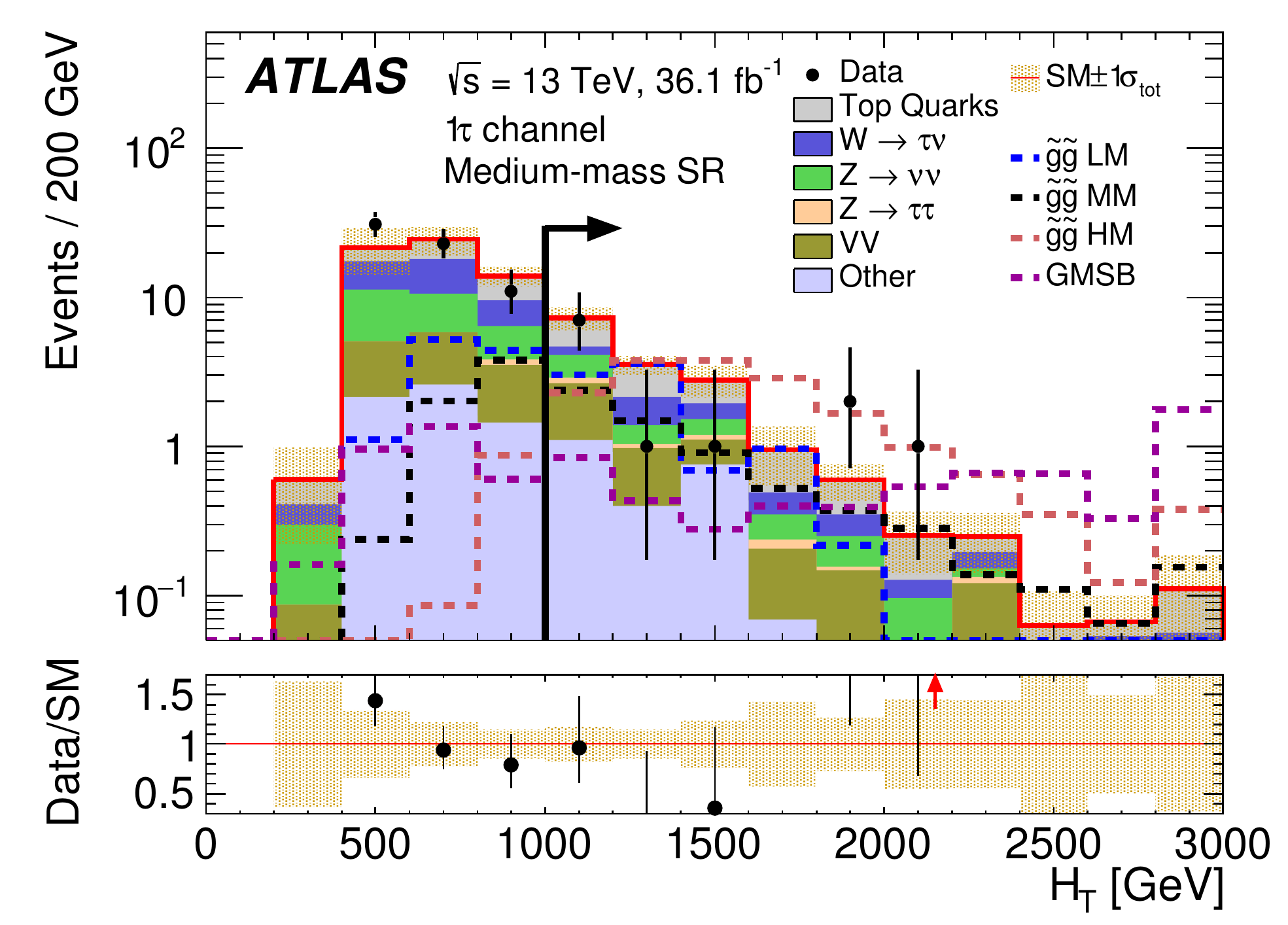}\label{fig:onetau:n_1:ht}}
\caption{Distributions of kinematic variables in extended SR selections of the \onetau channel after the fit: (a) $\tau$-lepton transverse mass \mTtau in the compressed SR without the $\mTtau > \SI{80}{\giga\electronvolt}$ requirement and (b) scalar sum of $\tau$-lepton and jet transverse momenta \HT in the medium-mass SR without the $\HT > \SI{1000}{\giga\electronvolt}$ requirement.
The contribution labeled as ``Other'' includes multijet events and the $V$+jets processes not explicitly listed in the legend.
The last bin of each distribution includes overflow events. The total uncertainty in the background prediction is shown as a shaded band. Arrows in the Data/SM ratio indicate bins where the entry is outside the plotted range. The signal region is indicated by the arrow in the upper pane. Signal predictions are overlaid for several benchmark models. 
For the simplified model, LM, MM and HM refer to low, medium and high mass-splitting scenarios, with $(m_{\tilde g},m_{\tilde\chi_1^0})$ set to $(1065,825)$~\gev, $(1625,905)$~\gev, and $(1705,345)$~\gev, respectively. 
The GMSB benchmark model corresponds to $\Lambda=\SI{120}{\tera\electronvolt}$ and $\tan\beta=40$.}
\label{fig:onetau:n_1}
\end{figure}

\begin{figure}
\centering
\subfigure[]{\includegraphics[width=0.50\textwidth]{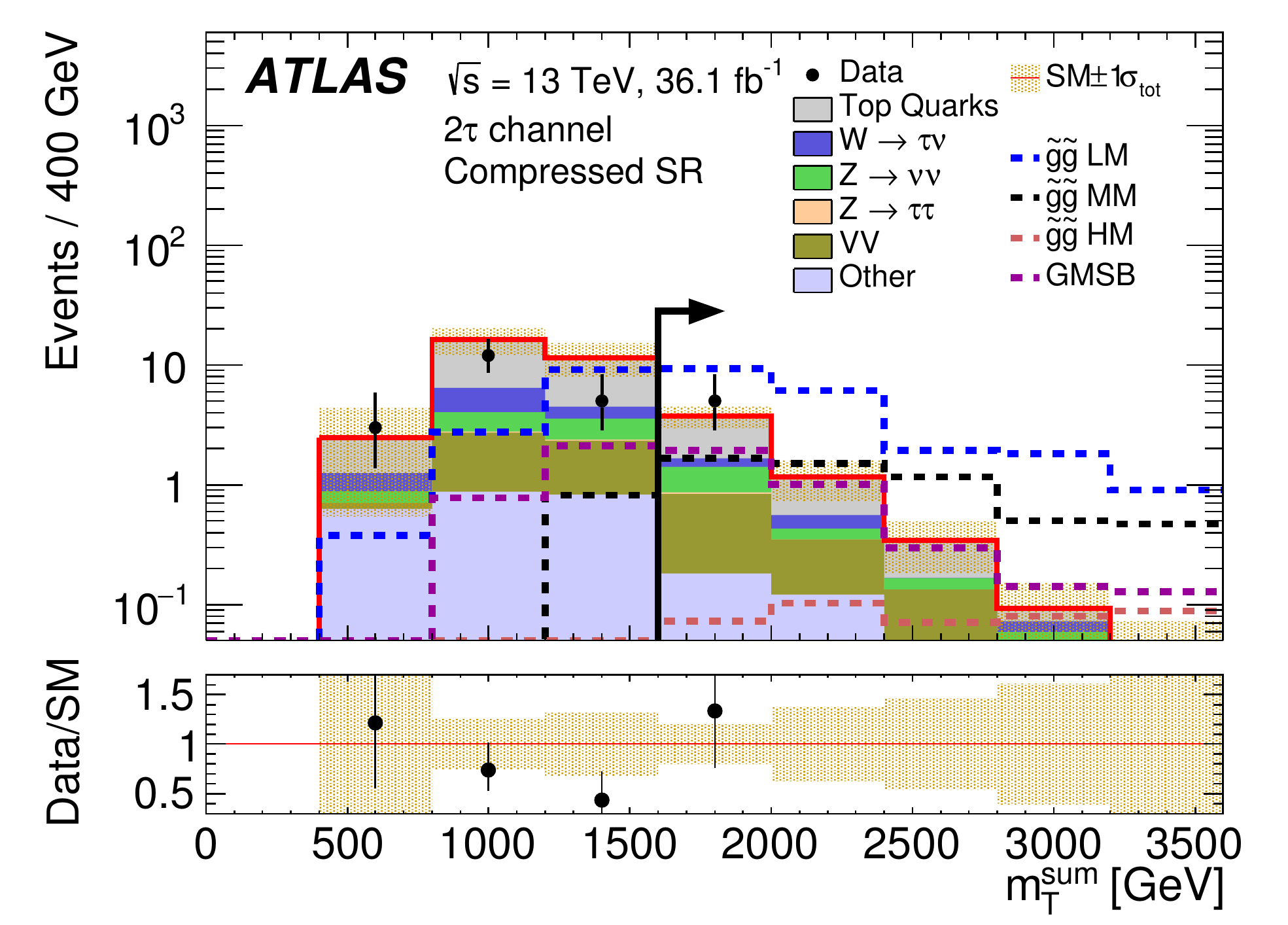}\label{fig:twotau:n_1:summttaujet}}
\hspace*{-1ex}\subfigure[]{\includegraphics[width=0.50\textwidth]{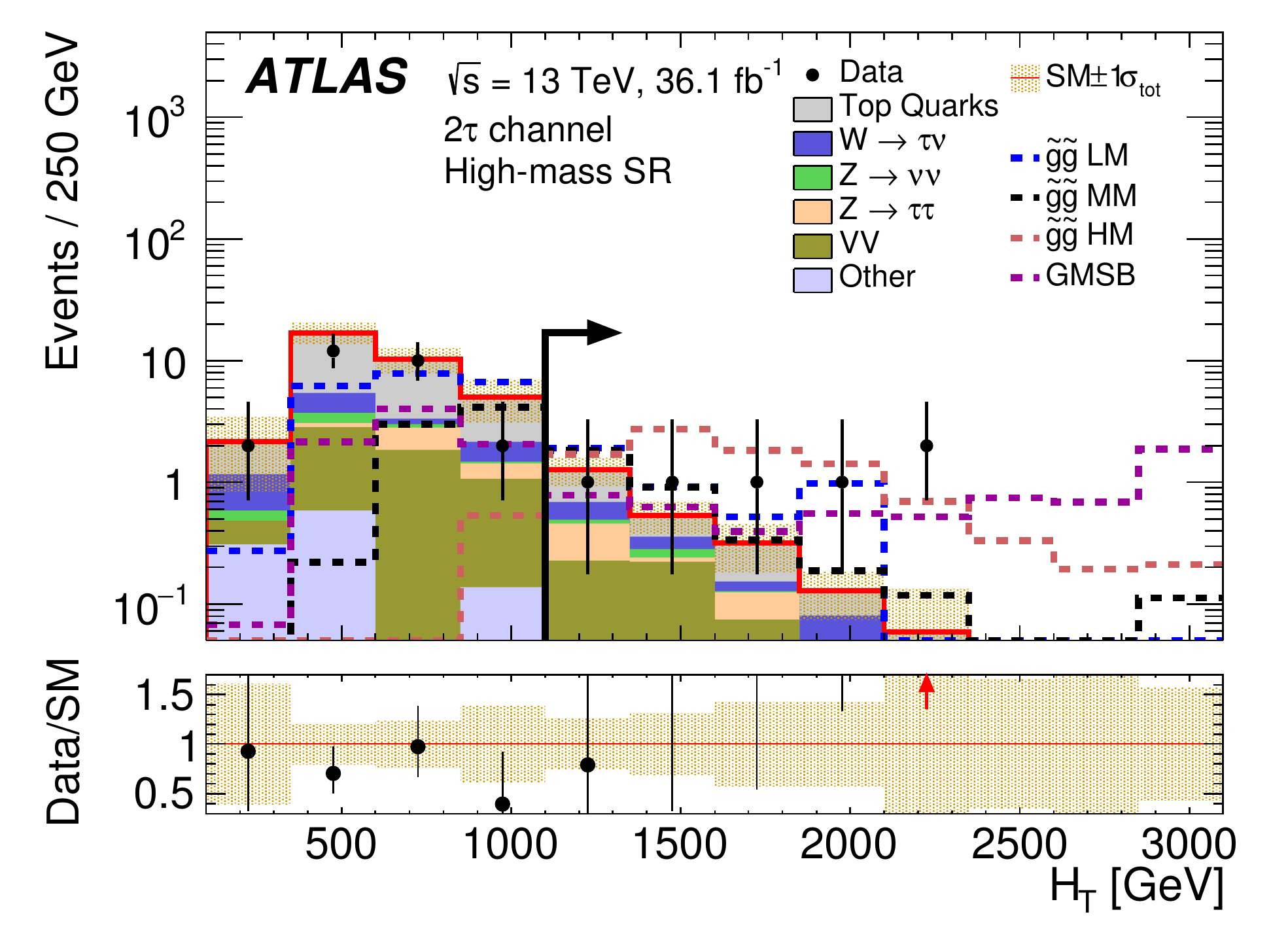}\label{fig:twotau:n_1:ht_hm}}
\hspace*{-1ex}\subfigure[]{\includegraphics[width=0.50\textwidth]{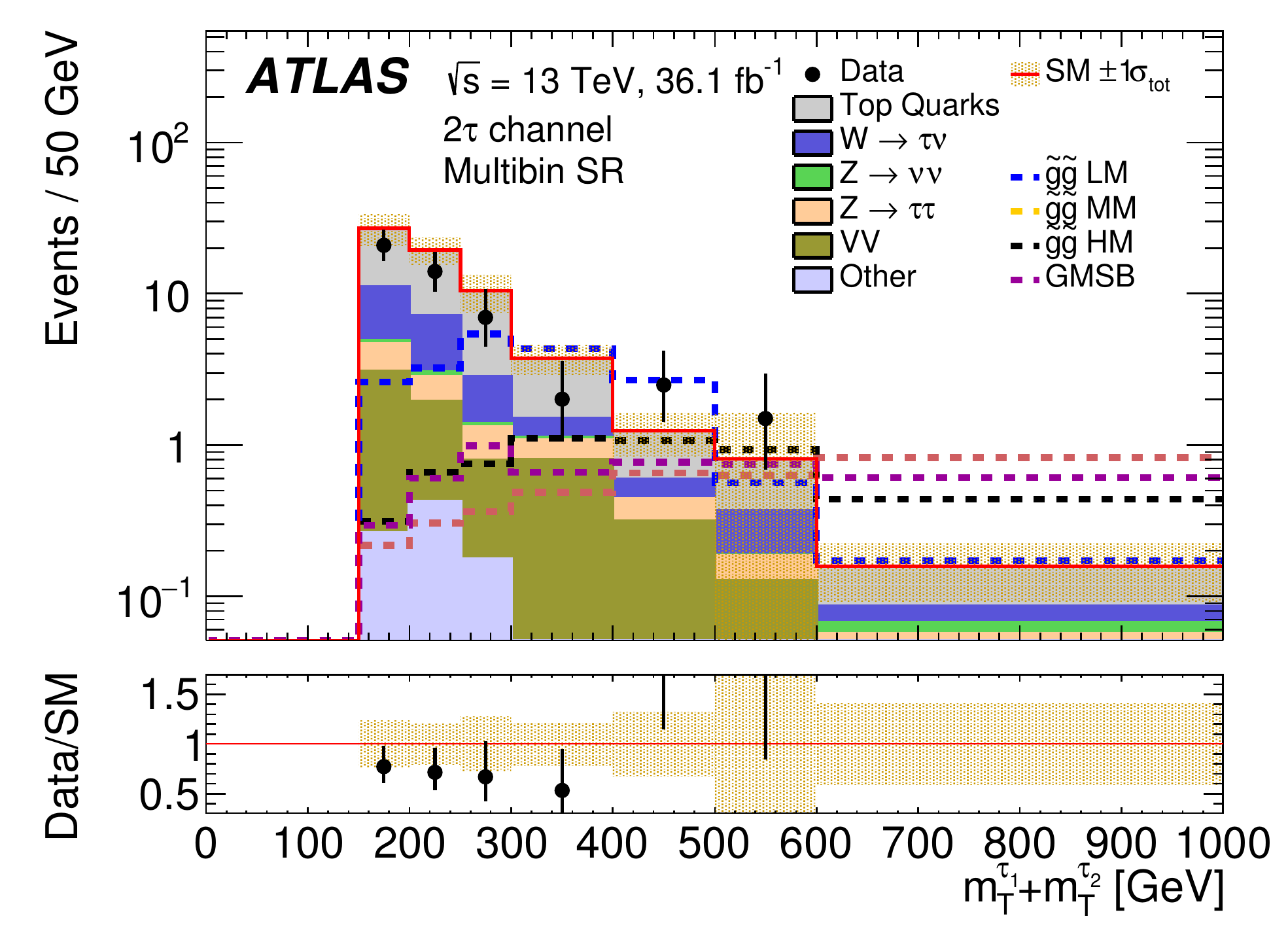}\label{fig:twotau:n_1:summt}}
\hspace*{-1ex}\subfigure[]{\includegraphics[width=0.50\textwidth]{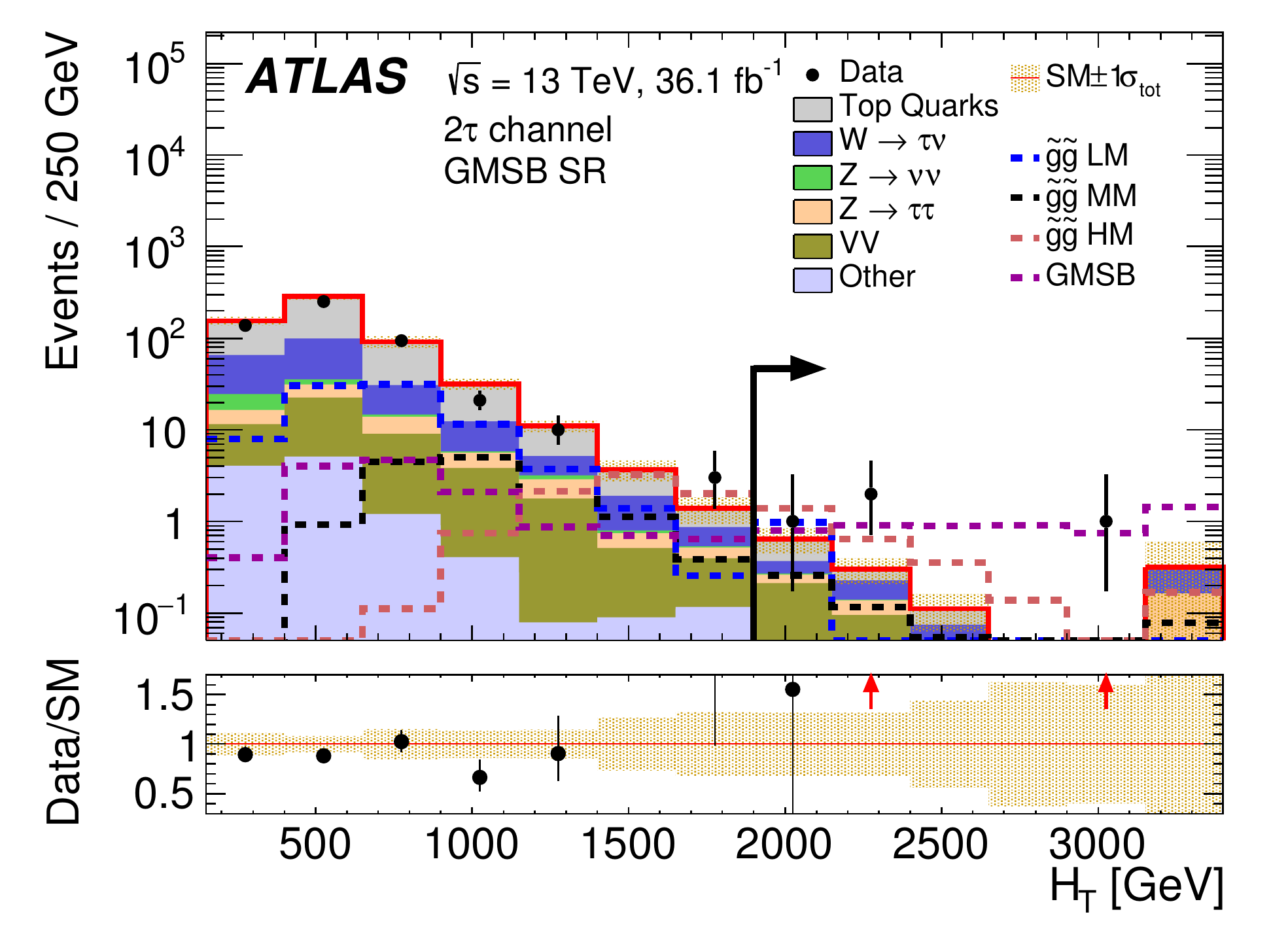}\label{fig:onetau:n_1:ht_gmsb}}
\caption{Distributions of kinematic variables in extended SR selections of the \twotau channel after the fit: (a) sum of transverse masses of $\tau$-leptons and jets $\mT^\mathrm{sum}$ in the compressed SR without the $\mT^\mathrm{sum} > \SI{1600}{\giga\electronvolt}$ requirement, (b) scalar sum of transverse momenta of $\tau$-leptons and jets \HT in the high-mass SR without the $\HT > \SI{1100}{\giga\electronvolt}$ requirement, (c) sum of transverse masses of $\tau$-leptons \mTTTkerned in the multibin SR, and (d) \HT in the GMSB SR without the $\HT > \SI{1900}{\giga\electronvolt}$ requirement.
The contribution labeled as ``Other'' includes multijet events and the $V$+jets processes not explicitly listed in the legend.
The last bin of each distribution includes overflow events. The total uncertainty in the background prediction is shown as a shaded band. Arrows in the Data/SM ratio indicate bins where the entry is outside the plotted range. The signal region is indicated by the arrow in the upper pane. Signal predictions are overlaid for several benchmark models. 
For the simplified model, LM, MM and HM refer to low, medium and high mass-splitting scenarios, with $(m_{\tilde g},m_{\tilde\chi_1^0})$ set to $(1065,825)$~\gev, $(1625,905)$~\gev, and $(1705,345)$~\gev, respectively. 
The GMSB benchmark model corresponds to $\Lambda=\SI{120}{\tera\electronvolt}$ and $\tan\beta=40$.}
\label{fig:twotau:n_1}
\end{figure}

The one-sided profile-likelihood-ratio test statistic is used to assess the probability that the observed data is compatible with the background-only and signal-plus-background hypotheses.
Systematic uncertainties are included in the likelihood function as nuisance parameters with Gaussian probability densities.
Following the standards used for LHC analyses, $p$-values are computed according to the $\mathrm{CL_s}$ prescription~\cite{0954-3899-28-10-313} using {\sc HistFitter}~\cite{Baak:2014wma}.

Model-independent upper limits on the event yields are calculated for each SR except the multibin SR, assuming no signal contribution in the CRs.
No such interpretation can be made for the multibin SR, as the relative signal contribution in each bin of the \mTTT distribution is model-dependent.
The results are derived using profile-likelihood-ratio distributions obtained from pseudo-experiments.
Upper limits on signal yields are converted into limits on the visible cross section ($\sigma_\mathrm{vis}$) of BSM processes by dividing by the integrated luminosity of the data.
The visible cross section is defined as the product of production cross section, acceptance, and selection efficiency. 
Results are summarized at the bottom of Tables~\ref{tab:results_mutual:yields:SR1T} and \ref{tab:results_mutual:yields:SR2T}.
The observed upper limits on the visible cross section range from \SI{0.18}{\femto\barn} for the compressed SR of the \twotau channel to \SI{1.37}{\femto\barn} for the compressed SR of the \onetau channel.

\begin{table}
\caption{Number of observed events and predicted background yields in the two signal regions of the \onetau channel. 
The background prediction is scaled using normalization factors derived in the control regions. The numbers in brackets give the background prediction before application of the fitted normalization factors. 
All systematic and statistical uncertainties are included in the quoted uncertainties.
The bottom part of the table shows the observed and expected model-independent upper limits at 95\% CL on the number of signal events $S_{\rm obs}^{95}$  and $ S_{\rm exp}^{95}  $, respectively, 
the corresponding observed upper limit on the visible cross section $\langle{\rm \sigma_{\rm vis}}\rangle_{\rm obs}^{95}$, the confidence level observed for the background-only hypothesis $\mathrm{CL}_{\mathrm{b}}$, 
the $p_0$-value, and corresponding significance Z. If the number of observed events is smaller than the expected background yield, the $p_0$-value is set to $0.5$, corresponding to a significance of $0.0$ standard deviations.}
\begin{center}
\setlength{\tabcolsep}{0.0pc}
{\footnotesize
\begin{tabular}{l>{\hspace{1.5cm}}r>{\hspace{.5cm}}r@{$\pm$}l>{\hspace{1.5cm}}r>{\hspace{.5cm}}r@{$\pm$}l}
\toprule
		{\bf \onetau channel}           & \multicolumn{3}{r}{Compressed SR}            & \multicolumn{3}{r}{Medium-mass SR}              \\[-0.05cm]
\midrule
		Data          && \multicolumn{2}{c}{$286$}              && \multicolumn{2}{c}{$12$}                    \\
\midrule
		Total background         &  [$290$] &$320 $&$ 32$         & [$15.2$]& $15.9 $&$ 3.0$              \\
\midrule
		Top quarks         & [$66$] &$77 $&$ 21$          & [$5.2$] &$5.8 $&$ 1.6$              \\
		$W(\tau\nu)$+jets         &  [$57$] &$51 $&$ 18$         & [$2.4$] &$2.2 $&$ 1.7$              \\
		$Z(\nu\nu)$+jets         &  [$77$] &$110 $&$ 24$         & [$1.5$] &$2.2 $&$ 0.5$              \\
		Other $V$+jets         &  [$52$] &$45 $&$ 10$         & [$1.9$] &$1.7 $&$ 0.4$              \\
		Diboson         &  [$28$] &$28 $&$ 5$         & [$3.0$] &$3.0 $&$ 0.6$              \\
		Multijet         & [$10.0$] &$9.2 $&$ 1.2$          & [$1.24$] &$1.14 $&$ 0.14$              \\
\midrule
\midrule
		$S_{\rm obs}^{95}$ ($S_{\rm exp}^{95}$)  &\multicolumn{3}{r}{$49.5$ ($ { 64.3 }^{ +24.1 }_{ -14.9 }$)}  &\multicolumn{3}{r}{$7.7$ ($ { 10.0 }^{ +4.3 }_{ -2.7 }$)}\\
		$\langle{\rm \sigma_{\rm vis}}\rangle_{\rm obs}^{95}$[fb] &&\multicolumn{2}{r}{ $1.37$} && \multicolumn{2}{r}{$0.21$} \\
		$\mathrm{CL}_{\mathrm{b}}$ &&\multicolumn{2}{r}{ $0.18$} && \multicolumn{2}{r}{$0.24$}\\
		$p_0\; (\mathrm{Z})$ &&\multicolumn{2}{r}{$0.5\; (0.0)$} && \multicolumn{2}{r}{$0.5\; (0.0)$}\\
\bottomrule
\end{tabular}
}
\end{center}
\label{tab:results_mutual:yields:SR1T}
\end{table}

\begin{table}
\caption{Number of observed events and predicted background yields in the three signal regions of the \twotau channel. 
The background prediction is scaled using normalization factors derived in the control regions. The numbers in brackets give the background prediction before application of the fitted normalization factors.
All systematic and statistical uncertainties are included in the quoted uncertainties.
The bottom part of the table shows the observed and expected model-independent upper limits at 95\% CL on the number of signal events $S_{\rm obs}^{95}$  and $ S_{\rm exp}^{95}  $, respectively, 
the corresponding observed upper limit on the visible cross section $\langle{\rm \sigma_{\rm vis}}\rangle_{\rm obs}^{95}$, the confidence level observed for the background-only hypothesis $\mathrm{CL}_{\mathrm{b}}$, 
the $p_0$-value, and corresponding significance (Z). If the number of observed events is smaller than the expected background yield, the $p_0$-value is set to $0.5$, corresponding to a significance of $0.0$ standard deviations.}
\begin{center}
\setlength{\tabcolsep}{0.0pc}
{\footnotesize
\begin{tabular}{l>{\hspace{1cm}}r>{\hspace{.5cm}}r@{$\pm$}l>{\hspace{1cm}}r>{\hspace{.5cm}}r@{$\pm$}l>{\hspace{1cm}}r>{\hspace{.5cm}}r@{$\pm$}l}
\toprule
{\bf \twotau channel}           & \multicolumn{3}{r}{Compressed SR}            & \multicolumn{3}{r}{High-mass SR}            & \multicolumn{3}{r}{GMSB SR}              \\[-0.05cm]
\midrule
Data          &&    \multicolumn{2}{c}{$5$}           && \multicolumn{2}{c}{$6$}              && \multicolumn{2}{c}{$4$}                  \\
\midrule
Total background         & [$4.7$]& $5.4 $&$ 1.9$          & [$2.3$]& $2.3 $&$ 0.7$          & [$1.5$]& $1.4 $&$ 0.5$              \\
\midrule
        Top quarks         &  [$2.3$]& $2.9 $&$ 1.7$         & [$0.9$]& $1.0 $&$ 0.5$          & [$0.34$]& $0.39 $&$ 0.23$             \\
        $W(\tau\nu)$+jets         &  [$0.5$]& \multicolumn{2}{l}{\hspace{28.25pt}$0.4_{-0.4}^{+0.5}$}         & [$0.4$]& $0.4 $&$ 0.4$          & [$0.4$]& $0.4 $&$ 0.4$              \\
        $Z(\tau\tau)$+jets         & [$0.035$]& $0.030 $&$ 0.011$          & [$0.37$]& $0.32 $&$ 0.11$          & [$0.33$]& $0.28 $&$ 0.10$              \\
        $Z(\nu\nu)$+jets         & [$0.47$]& $0.67 $&$ 0.35$          & [$0.065$]& $0.093 $&$ 0.028$          & [$0.008$]& $0.011 $&$ 0.007$              \\
        Other $V$+jets         &  [$0.32$]& $0.30 $&$ 0.08$         & [$0.019$]& $0.015 $&$ 0.012$          & [$<0.01$]& \multicolumn{2}{c}{$<0.01$}              \\
        Diboson         &   [$1.06$]& $1.05 $&$ 0.25$        & [$0.56$]& $0.56 $&$ 0.15$          & [$0.29$]& $0.29 $&$ 0.08$              \\
        Multijet         &  [$0.0261$]& $0.0241 $&$ 0.0031$         & [$0.0131$]& $0.0121 $&$ 0.0015$          & [$0.065$]& $0.060 $&$ 0.008$              \\
\midrule
\midrule
$S_{\rm obs}^{95}$ ($S_{\rm exp}^{95}$) & \multicolumn{3}{r}{$6.7$ ($ { 6.7 }^{ +2.8 }_{ -1.5 }$)} & \multicolumn{3}{r}{$9.0$ ($ { 5.0 }^{ +1.9 }_{ -1.3 }$)} & \multicolumn{3}{r}{$7.3$ ($ { 4.4 }^{ +1.5 }_{ -0.9 }$)}\\
$\langle{\rm \sigma_{\rm vis}}\rangle_{\rm obs}^{95}$[fb] & &\multicolumn{2}{r}{$0.18$} & &\multicolumn{2}{r}{$0.25$} & &\multicolumn{2}{r}{$0.20$}\\
$\mathrm{CL}_{\mathrm{b}}$ & &\multicolumn{2}{r}{$0.50$} & &\multicolumn{2}{r}{$0.96$} & &\multicolumn{2}{r}{$0.95$}\\
$p_0\; (\mathrm{Z})$ & &\multicolumn{2}{r}{$0.5\; (0.0)$} & &\multicolumn{2}{r}{$0.03\; (1.83)$} & &\multicolumn{2}{r}{$0.05\; (1.68)$}\\

\bottomrule
\end{tabular}
}
\end{center}
\label{tab:results_mutual:yields:SR2T}
\end{table}

Limits are also set for the two SUSY models discussed in Section~\ref{sec:intro}.
Exclusion contours at the 95\%~CL are derived in the $(m_{\tilde g},m_{\tilde\chi_1^0})$ parameter space for the simplified model and in the $(\Lambda,\tan\beta)$ parameter space for the GMSB model.
In the case of model-dependent interpretations, the signal contribution in the control regions is included in the calculation of upper limits, and asymptotic properties of test-statistic distributions are used~\cite{Cowan:2010js}.
Results are shown in Figures~\ref{fig:contours:BestCLS} and \ref{fig:contours:GMSB_Combo}.
The solid line and the dashed line correspond to the observed and median expected limits, respectively.
The band shows the one-standard-deviation spread of the expected limits around the median, which originates from statistical and systematic uncertainties in the background and signal.
The theoretical uncertainty in the signal cross section is not included in the band. Its effect on the observed limits is shown separately as dotted lines.
For both SUSY models, the exclusion limits obtained with $36.1~\ifb$ of collision data at $\sqrt{s}=13 \tev$ significantly improve upon the previous ATLAS results~\cite{SUSY-2016-01} established with $3.2~\ifb$ of $13 \tev$ data.
Besides the increase in the integrated luminosity, the results benefit from an improved analysis and statistical treatment.
The \onetau and \twotau channels are now statistically combined in a global fit, while in the previous analysis, only the SR with the lowest expected $\mathrm{CL_s}$ value was considered for the simplified model, 
and only the \twotau GMSB SR was used for the GMSB interpretation.
In addition, the multibin SR of the \twotau channel provides increased sensitivity to gluino pair production over a large region of the parameter space.

Expected limits in the model parameter space are shown for each channel, to illustrate their complementarity and the gain in sensitivity achieved with their combination.
The green dash-dotted line corresponds to a fit that includes all CRs and the two SRs of the \onetau channel.
For the \twotau channel, in the case of the simplified model, the magenta dash-dotted line corresponds to the best expected exclusion from fits that include either the \twotau multibin SR or the combination of the \twotau compressed and high-mass SRs.
In the GMSB model, the \twotau combination is based on the \twotau GMSB and compressed SRs.
In the simplified model, the \onetau and \twotau channels have similar sensitivity at high gluino and low LSP masses. 
For high LSP masses, the combination is dominated by the \twotau channel, while in the region with a low mass difference between the gluino and the LSP, the \onetau channel drives the exclusion.
In the GMSB interpretation, the more stringent limits at high values of $\tan\beta$ are explained by the nature of the NLSP, which is the lightest $\tau$-slepton in this region. 
For lower values of $\tan\beta$, the $\PSgti$ is nearly mass-degenerate with $\PSe_{\text{R}}$ and $\PSgm_{\text{R}}$, leading to fewer $\tau$-leptons in squark and gluino decays, and reduced sensitivity of the \twotau GMSB SR.
The weaker exclusion at low $\tan\beta$ is mitigated by the SRs from the \onetau channel and the compressed SR of the \twotau channel. For high $\Lambda$, the sensitivity is limited by the strong-production cross section. While the analysis is mainly sensitive to squark and gluino production, the total GMSB production cross section for high $\Lambda$ is dominated by electroweak production modes.

\begin{figure}[tb]
 \centering
  {\includegraphics[width=0.88\textwidth]{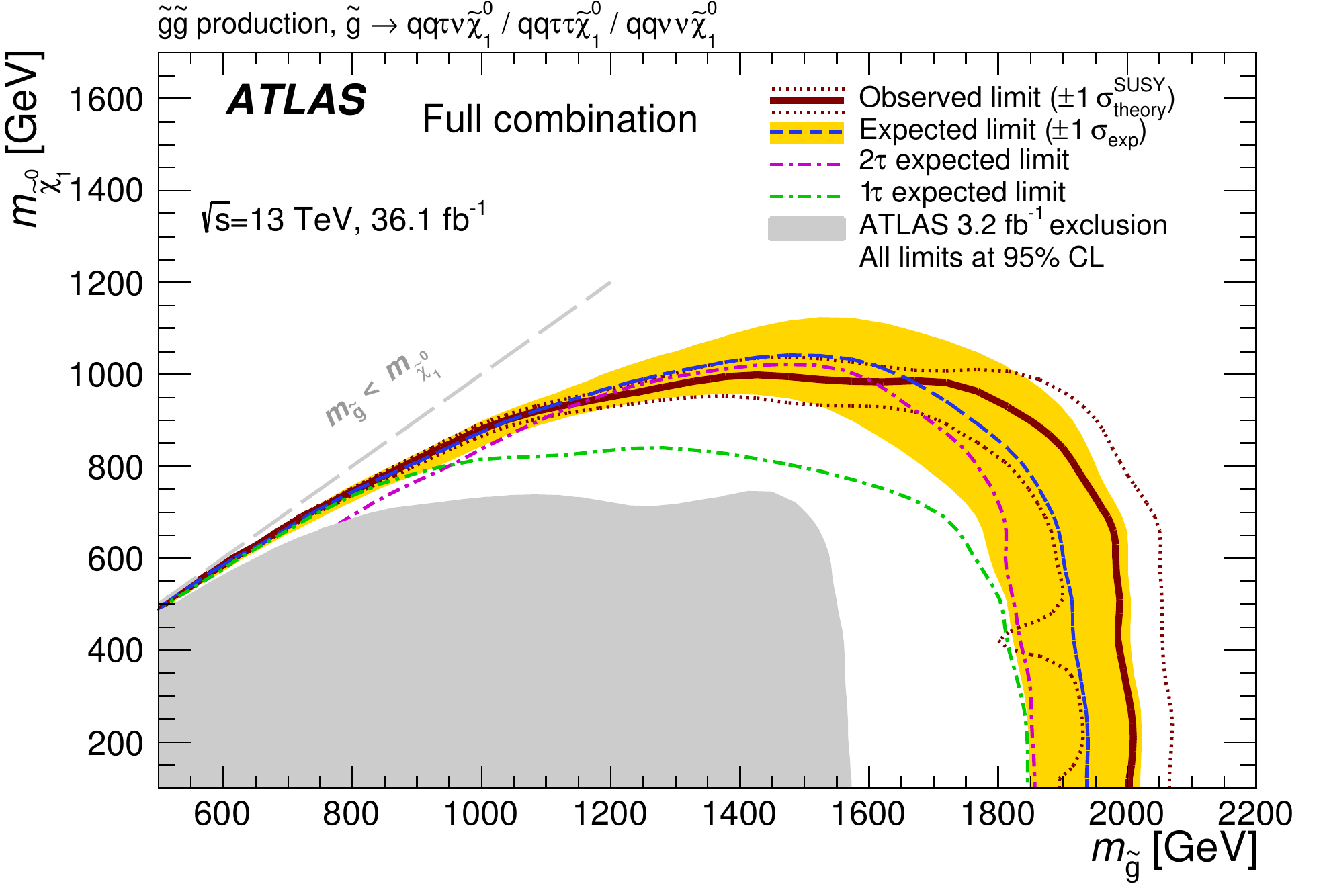}}
  \caption{Exclusion contours at the 95\% confidence level as a function of the LSP mass $m_{\tilde\chi_1^0}$ 
and gluino mass $m_{\tilde g}$ for the simplified model of gluino pair production.
The solid line and the dashed line correspond to the observed and median expected limits, respectively, for the combination of the \onetau and \twotau channels.
The band shows the one-standard-deviation spread of expected limits around the median.
The effect of the signal cross-section uncertainty on the observed limits is shown as dotted lines.
The inward fluctuation of the $-1\sigma$ line originates from the method employed to perform the combination.
The previous ATLAS result~\cite{SUSY-2016-01} obtained with $3.2~\ifb$ of $13 \tev$ data is shown as the filled area in the bottom left.}
   \label{fig:contours:BestCLS}
\end{figure}

\clearpage

\begin{figure}[tb]
 \centering
  {\includegraphics[width=0.88\textwidth]{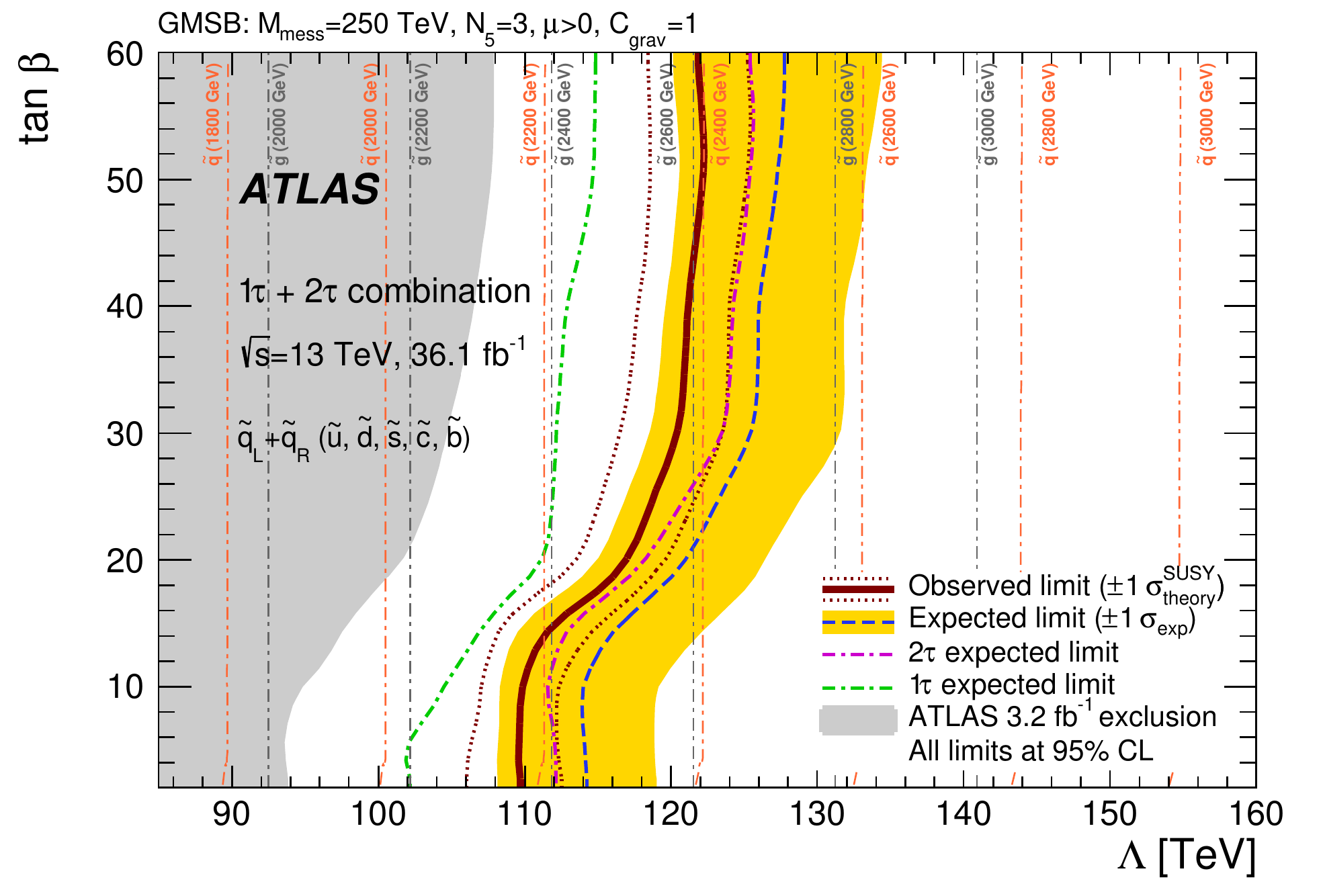}}
  \caption{Exclusion contours at the 95\% confidence level as a function of $\tan\beta$ and the SUSY-breaking 
mass scale $\Lambda$ for the gauge-mediated supersymmetry-breaking model.
The solid line and the dashed line correspond to the observed and median expected limits, respectively, for the combination of the \onetau and \twotau channels.
The band shows the one-standard-deviation spread of expected limits around the median.
The effect of the signal cross-section uncertainty on the observed limits is shown as dotted lines.
 The gray and orange dash-dotted lines indicate the masses of gluinos and mass-degenerate squarks, respectively.
 The previous ATLAS result~\cite{SUSY-2016-01} obtained with $3.2~\ifb$ of $13 \tev$ data is shown as the filled area on the left.
}
   \label{fig:contours:GMSB_Combo}
\end{figure}

\FloatBarrier

\section{Conclusion}
\label{sec:conclusion}

A search for squarks and gluinos in events with jets, hadronically decaying $\tau$-leptons, and missing transverse momentum is performed using $pp$ collision data at $\sqrt{s} = 13$~\tev\ recorded by the ATLAS detector at the LHC in 2015 and 2016, corresponding to an integrated luminosity of $36.1~\ifb$. 
Two channels with exactly one or at least two $\tau$-leptons are considered, and their results are statistically combined. 
The observed data are consistent with background expectations from the Standard Model.
Upper limits are set at 95\% confidence level on the number of events that could be produced by processes beyond the Standard Model.
Results are also interpreted in the framework of a simplified model of gluino pairs decaying into $\tau$-leptons via $\tau$-sleptons, and a minimal model of gauge-mediated supersymmetry breaking with the lighter $\tau$-slepton as the NLSP at large $\tan\beta$. At 95\% CL in the simplified model, gluino masses up to 2000~\gev\ are excluded for low LSP masses, and LSP masses up to 1000~\gev\ are excluded for gluino masses around 1400~\gev.
In the GMSB model, values of the SUSY-breaking scale $\Lambda$ below 110~\tev\ are excluded at 95\% CL for all values of $\tan\beta$ in the range $2\leq\tan\beta\leq60$, while a stronger limit of 120~\tev\ is achieved for $\tan\beta>30$.

\section*{Acknowledgments}


We thank CERN for the very successful operation of the LHC, as well as the
support staff from our institutions without whom ATLAS could not be
operated efficiently.

We acknowledge the support of ANPCyT, Argentina; YerPhI, Armenia; ARC, Australia; BMWFW and FWF, Austria; ANAS, Azerbaijan; SSTC, Belarus; CNPq and FAPESP, Brazil; NSERC, NRC and CFI, Canada; CERN; CONICYT, Chile; CAS, MOST and NSFC, China; COLCIENCIAS, Colombia; MSMT CR, MPO CR and VSC CR, Czech Republic; DNRF and DNSRC, Denmark; IN2P3-CNRS, CEA-DRF/IRFU, France; SRNSFG, Georgia; BMBF, HGF, and MPG, Germany; GSRT, Greece; RGC, Hong Kong SAR, China; ISF, I-CORE and Benoziyo Center, Israel; INFN, Italy; MEXT and JSPS, Japan; CNRST, Morocco; NWO, Netherlands; RCN, Norway; MNiSW and NCN, Poland; FCT, Portugal; MNE/IFA, Romania; MES of Russia and NRC KI, Russian Federation; JINR; MESTD, Serbia; MSSR, Slovakia; ARRS and MIZ\v{S}, Slovenia; DST/NRF, South Africa; MINECO, Spain; SRC and Wallenberg Foundation, Sweden; SERI, SNSF and Cantons of Bern and Geneva, Switzerland; MOST, Taiwan; TAEK, Turkey; STFC, United Kingdom; DOE and NSF, United States of America. In addition, individual groups and members have received support from BCKDF, the Canada Council, CANARIE, CRC, Compute Canada, FQRNT, and the Ontario Innovation Trust, Canada; EPLANET, ERC, ERDF, FP7, Horizon 2020 and Marie Sk{\l}odowska-Curie Actions, European Union; Investissements d'Avenir Labex and Idex, ANR, R{\'e}gion Auvergne and Fondation Partager le Savoir, France; DFG and AvH Foundation, Germany; Herakleitos, Thales and Aristeia programmes co-financed by EU-ESF and the Greek NSRF; BSF, GIF and Minerva, Israel; BRF, Norway; CERCA Programme Generalitat de Catalunya, Generalitat Valenciana, Spain; the Royal Society and Leverhulme Trust, United Kingdom.

The crucial computing support from all WLCG partners is acknowledged gratefully, in particular from CERN, the ATLAS Tier-1 facilities at TRIUMF (Canada), NDGF (Denmark, Norway, Sweden), CC-IN2P3 (France), KIT/GridKA (Germany), INFN-CNAF (Italy), NL-T1 (Netherlands), PIC (Spain), ASGC (Taiwan), RAL (UK) and BNL (USA), the Tier-2 facilities worldwide and large non-WLCG resource providers. Major contributors of computing resources are listed in Ref.~\cite{ATL-GEN-PUB-2016-002}.

\clearpage
\printbibliography

\clearpage 
 
\begin{flushleft}
{\Large The ATLAS Collaboration}

\bigskip

M.~Aaboud$^\textrm{\scriptsize 34d}$,    
G.~Aad$^\textrm{\scriptsize 99}$,    
B.~Abbott$^\textrm{\scriptsize 124}$,    
O.~Abdinov$^\textrm{\scriptsize 13,*}$,    
B.~Abeloos$^\textrm{\scriptsize 128}$,    
D.K.~Abhayasinghe$^\textrm{\scriptsize 91}$,    
S.H.~Abidi$^\textrm{\scriptsize 164}$,    
O.S.~AbouZeid$^\textrm{\scriptsize 143}$,    
N.L.~Abraham$^\textrm{\scriptsize 153}$,    
H.~Abramowicz$^\textrm{\scriptsize 158}$
% ...
\bigskip
\\

% ...
$^{13}$Institute of Physics, Azerbaijan Academy of Sciences, Baku; Azerbaijan.\\
% ...
$^{34}$$^{(a)}$Facult\'e des Sciences Ain Chock, R\'eseau Universitaire de Physique des Hautes Energies - Universit\'e Hassan II, Casablanca;$^{(b)}$Centre National de l'Energie des Sciences Techniques Nucleaires (CNESTEN), Rabat;$^{(c)}$Facult\'e des Sciences Semlalia, Universit\'e Cadi Ayyad, LPHEA-Marrakech;$^{(d)}$Facult\'e des Sciences, Universit\'e Mohamed Premier and LPTPM, Oujda;$^{(e)}$Facult\'e des sciences, Universit\'e Mohammed V, Rabat; Morocco.\\
% ...
$^{91}$Department of Physics, Royal Holloway University of London, Egham; United Kingdom.\\
% ...
$^{99}$CPPM, Aix-Marseille Universit\'e, CNRS/IN2P3, Marseille; France.\\
% ...
$^{124}$Homer L. Dodge Department of Physics and Astronomy, University of Oklahoma, Norman OK; United States of America.\\
% ...
$^{128}$LAL, Universit\'e Paris-Sud, CNRS/IN2P3, Universit\'e Paris-Saclay, Orsay; France.\\
% ...
$^{143}$Santa Cruz Institute for Particle Physics, University of California Santa Cruz, Santa Cruz CA; United States of America.\\
% ...
$^{153}$Department of Physics and Astronomy, University of Sussex, Brighton; United Kingdom.\\
% ...
$^{158}$Raymond and Beverly Sackler School of Physics and Astronomy, Tel Aviv University, Tel Aviv; Israel.\\
% ...
$^{164}$Department of Physics, University of Toronto, Toronto ON; Canada.\\
% ...
$^{a}$ Also at  Department of Physics, University of Malaya, Kuala Lumpur; Malaysia.\\
$^{b}$ Also at Borough of Manhattan Community College, City University of New York, NY; United States of America.\\
$^{c}$ Also at California State University, East Bay; United States of America.\\
$^{d}$ Also at Centre for High Performance Computing, CSIR Campus, Rosebank, Cape Town; South Africa.\\
$^{e}$ Also at CERN, Geneva; Switzerland.\\
% ...
$^{*}$ Deceased

\end{flushleft}

 

\end{document}